\documentclass[pra,twocolumn,showpacs,superscriptaddress]{revtex4-2}

\usepackage{epsf,epsfig}
\usepackage[psamsfonts]{amssymb}
\usepackage{amsmath}
\usepackage{bm}
\usepackage{natbib}
\usepackage{graphicx}
\usepackage{color}
\usepackage{mathrsfs}
\RequirePackage{chapterbib}

\usepackage{tabularx}
\usepackage[labelformat=simple]{subcaption}

\usepackage{graphicx}

\begin{document}
\title{Nonlinear optical effects on the atom-field interaction based on the nonlinear coherent states approach}
\author{Mojgan Momeni Demneh}
\email{mojganmomeni@outlook.com}
\affiliation{Department of Physics, University of Isfahan, Hezar Jerib 81746-73441, Isfahan, Iran}
\author{ Ali Mahdifar}
\email{a.mahdifar@sci.ui.ac.ir}
\affiliation{Department of Physics, University of Isfahan, Hezar Jerib 81746-73441, Isfahan, Iran}
\affiliation{Quantum Optics Group, Department of Physics, University of Isfahan, Hezar Jerib 81746-73441, Isfahan, Iran}
\author{Rasoul Roknizadeh}
\email{rokni@sci.ui.ac.ir }
\affiliation{Department of Physics, University of Isfahan, Hezar Jerib 81746-73441, Isfahan, Iran}
\affiliation{Quantum Optics Group, Department of Physics, University of Isfahan, Hezar Jerib 81746-73441, Isfahan, Iran}
\begin{abstract}
In this paper, to study the effects of a nonlinear medium on the atom-field interaction, we use the nonlinear coherent states approach. For this purpose, we choose the two-mode cross-Kerr as the our nonlinear optical phenomena and with the use of it's algebra, we show that it can be described equivalently by a deformed oscillator algebra and also, by a deformed $su(2)$ algebra. Then, we construct the associated coherent states and investigate their statistical properties. After that, as an example of applicability of the constructed two-mode nonlinear coherent states, we investigate the nonlinear effects of the medium on the dynamics of atom-field interaction within the framework of the coherent states. By using the time-dependent Schrödinger equation, we first obtain the atom-field state and then study the effect of the nonlinear medium on the occupation probabilities of the atomic levels. In the following, we consider the relation between the revival time of the atomic occupation probabilities and the nonlinear parameter of the medium. Then, to study the nonlinear effects on the dynamical properties of the cavity field, we consider photon distribution, correlation function, Mandel parameters of the field, the von Neumann entropy and the squeezing. Particularly, the nonlinearity of the media on the nonclassical properties of two modes is clarified.
\end{abstract}
%
\maketitle
\section{Introduction}
In nonlinear optics, the response of a nonlinear media to a high intensity light is considered. This nonlinear response which is typically observed only at very high intensities light, can give rise to a host of effects, including frequency conversion and amplification, and have fundamental interest and practical importance which has long received much attention~\cite{lindsay,kielich1981, shen1984principles, boyd14}. However, after that the high-intensity light of a laser made possible the first experimental demonstration of second harmonic generation (frequency up-conversion) in 1961~\cite{franken1961generation}, many more nonlinear-optics effects were demonstrated using a variety of nonlinear media. As is known, the Kerr and cross-Kerr effects are of the most common effects in nonlinear optics. In a medium of these types, the refractive index of the media consists of two terms, the first is constant, and is just the usual linear index of refraction, while the second is proportional to the intensity of the field~\cite{kockum2017deterministic}.

The corresponding effects in nonlinear optics are studied by interaction of classical electromagnetic fields with dielectric media. While, some of most important quantum mechanically interesting phenomena can be observed by quantization of electromagnetic field in nonlinear media~\cite{kockum2017deterministic,chang2014quantum,stassi2017quantum, kheirandish2011finite}. Accordingly, by studying quantum nonlinear optics (QNO), we may control the information processing systems via photon-by photon~\cite{chang2014quantum}. For example, QNO without photons~\cite{stassi2017quantum} and with single atoms and virtual photons~\cite{kockum2017deterministic} are the cases have been studied. Moreover, the realization of quantum nonlinear optics could improve the performance of the classical nonlinear devices~\cite{miller2010optical}. Furthermore, nonlinear switches activated by single photons could enable optical quantum information processing and communication~\cite{kimble2008quantum}, as well as other applications that rely on the generation and manipulation of non-classical light fields~\cite{muthukrishnan2004quantum}. Achieving strong interactions between individual photons has been a goal at the forefront of the quantum-optics research for several decades. These interactions constitute a fundamental tool towards the ultimate control of light fields ‘quantum by quantum’~\cite{firstenberg2013quantum}.

On the other hand, by studying the properties of photons emitted from a nonlinear medium, it is shown that these photons exhibit nonclassical properties, such as squeezing.  Also, as is known, by using a nonlinear optical process, it is possible to generate the squeezed light~\cite{wu1987squeezed, wu1986generation}. The second order nonlinear effect of optical parametric amplification is probably the most successful technique in the generation of the squeezed states of light~\cite{ast2015new}. In addition, the Kerr interaction creates a photon number squeezed state with sub-Poissonian statistics~\cite{silberhorn2001generation}. Furthermore, generation of twin photons inside a nonlinear crystal by utilizing the spontaneous parametric downconversion (SPDC) phenomenon has been extremely useful for studying fundamental aspects of quantum mechanics and physically implementing the quantum information protocols~\cite{saldanha2013energy, howell2004realization}.

On the other side, coherent states (CSs) of the harmonic oscillator~\cite{glauber1963coherent, glauber1963photon, glauber1963quantum} as well as generalized CSs associated with various algebra~\cite{perelomov2012generalized, klauder1985coherent, ali2000coherent} play an important role in various theoretical and experimental fields of modern physics, including quantum optics and quantum communication. Among the generalized CSs, f-deformed CSs or nonlinear CSs (NCSs)~\cite{katriel1994nonideal, shanta1994unified} have attracted much attention in recent years, mostly because they exhibit nonclassical properties, such as amplitude squeezing and quantum interference~\cite{vogel1995nonlinear, de1996nonlinear}. As mentioned in~\cite{roknizadeh2004construction}, up to now many quantum optical states such as q-deformed CSs~\cite{man1997f}, photon added CSs and photon-subtracted CSs~\cite{sivakumar1999photon, naderi2004new}, negative binomial states~\cite{liao2000superpositions}, and the center of mass motion of a trapped ion~\cite{vogel1995nonlinear, mahdifar2008coherent} have been considered as some kind of nonlinear CSs. These states are associated with nonlinear algebras and are defined as the eigenstates of the annihilation operator of an f-deformed oscillator~\cite{man1997f}. As is well known, the coherent states as an overcomplete set in Hilbert space of a quantum system, can represent the geometry of the physical background space~\cite{vogel1995nonlinear, mahdifar2008coherent}. In other hand, in some recent work it is shown the nonlinearty has some equivalent geometry~\cite{mousavi2016induced}.

The goal of this contribution is twofold. In the first part, our purpose is to study how the nonlinearity can be represented in related coherent states. In order to do so, we choose the quantum two-mode cross-Kerr as our nonlinear system and describe it as a nonlinear (f-deformed) two-dimensional oscillator. Also, by using a nonlinear Schwinger model, we show that it is possible to describe the above-mentioned system as a deformed $su(2)$. Then, we construct the corresponding CSs and investigate their quantum statistical properties. In the second part of the paper, we try to investigate the nonlinear effects of the medium on the dynamics of atom-field interaction within the framework of the nonlinear CSs. To achieve this goal, we consider the interaction of a two-mode field with a three-level atom in the present of the nonlinear medium (cross-Kerr). Then, we model this system as a three–level atom-field in free space and treat the nonlinear effects of the medium throughout the corresponding NCSs.  As the outputs of the motioned interaction, we investigate the effects of nonlinearity on the atomic occupation probabilities, the cross-correlation function, the Mandel parameter, the squeezing and the von Neumann entropy.

The paper is organized as follows. In section II, we briefly review the NCSs. In section III, we describe the algebra of a nonlinear optics phenomena (cross-Kerr effect) with the f-deformed oscillator algebra and then show that algebra of our nonlinear optical system can be identified as a new type of deformed $su(2)$ algebra.
The two-dimensional nonlinear CSs associated to our nonlinear optical system are obtained in section IV. Sections V is devoted to the investigation of the quantum statistical properties and also the entanglement between two-modes of the constructed two-mode constructed CSs. Particularly, the nonlinearity of the media on the nonclassical properties of two modes is clarified. In section VI, as the second part of manuscript, we study the interaction of a two-mode cross-Kerr nonlinear CSs (CK-NCSs) with a three-level atom. Sections VII and VIII are devoted to the effects nonlinearity of the media on the atomic occupation probabilities, the cross-correlation function and the Mandel parameter. In particular, the normal squeezing of the field quadratures and the quantum entanglement due to the atom–field interaction are evaluated. Finally, the summary and concluding remark are presented in section IX.
\section{Nonlinear coherent states}
The f-deformed quantum oscillator~\cite{man1997f} as a nonlinear oscillator with a specific kind of nonlinearity, is characterizes by the following deformed dynamical operators:
\begin{equation}
\begin{split}
\begin{array}{l}
\hat A = \hat af(\hat n) = f(\hat n + 1)\hat a,\\
{{\hat A}^\dag } = {f^\dag }(\hat n){{\hat a}^\dag } = {{\hat a}^\dag }{f^\dag }(\hat n + 1),\end{array}
\end{split}\tag{1}
\end{equation}
where $\hat a$, ${\hat a^\dag }$ and $\hat n = {\hat a^\dag }\hat a$ are the bosonic annihilation, creation and number operators, respectively, and $f(\hat n)$ is the deformation function. Ordinarily, the phase of $f$ is irrelevant and it is possible to choose $f$ to be a nonnegative real function, i.e., ${f^\dag }(\hat n) = f(\hat n)$. The commutator between deformed operators $\hat A$ and $\hat A^\dag $ is given by :
\begin{equation}
[\hat A,{\hat A^\dag }] = (\hat n + 1){f^2}(\hat n + 1) - \hat n{f^2}(\hat n).\tag{2}
\end{equation}
As is seen, the nature of the above deformed Weyl-Heisenberg algebra depends on the nonlinear deformation function $f(\hat n)$ and in the limiting case $f(\hat n) \to 1,$     will reduce to the conventional Weyl-Heisenberg algebra.
\section{f-deformed oscillator algebra’s description of a nonlinear optics phenomena}
In this section we show that a quantum nonlinear optical phenomena can be describe by the f-deformed oscillator approach. As is known, the most common phenomenon (cross-Kerr effect) to arise out of a ${\chi ^{\left( 3 \right)}}$ nonlinearity is that of intensity-dependent refractive index~\cite{kockum2017deterministic}.
Although the photons do not interact with each other through the vacuum space, some proposed approaches in quantum information processes are based on some form of interaction between photons, using the nonlinear optics. Among these nonlinear optical processes, the nonlinear Kerr interaction between cavity modes has been studied widely. Recent studies have shown that the cross-Kerr nonlinearity, or the so-called cross phase modulation (XPM), can be used for quantum information processes including the construction of nontrivial quantum gates~\cite{turchette1995measurement}, the preparation of entangled photon states~\cite{xiao2006quantum}, and quantum non-demolition measurement~\cite{barrett2005symmetry}. On the other hand, one of the most challenge for the nonlinear quantum optics is the realization of photon-photon interactions at the level of a few photons. For that, the XPM is often used to realize the interaction between two photons modes which coupled to a nonlinear atomic media. Also, to minimize the dispersion and the absorption induced by the decoherence of the atoms, schemes in N-type atoms have been proposed and realized~\cite{hu2011cross, schmidt1996giant, kang2003observation}.

The quantum mechanical description of two modes cross-Kerr is given by the following Hamiltonian:
\begin{equation}
 \hat H = \hbar {\omega _a}{\hat a^\dag }\hat a + \hbar {\omega _b}{\hat b^\dag }\hat b + \frac{{\hbar \kappa }}{2}{\hat a^\dag }{\hat b^\dag }\hat a\hat b, \tag{3}
\end{equation}
where $\hat a\left( {{{\hat a}^\dag }} \right)$ and $\hat b\left( {{{\hat b}^\dag }} \right)$ are the annihilation (creation) operators for the signal and pump modes, respectively. The first and second terms in the above equation describe the free-field Hamiltonian. Moreover, the last term which is proportional to $\kappa $, describes the cross-Kerr interaction of the system. It is worth noting that the parameter of nonlinearity $\kappa $ is proportional to ${\chi ^{\left( 3 \right)}}$~\cite{hillery2009introduction, imamoglu1997strongly}.

According to Eq. (3), it can be easily seen that the energy eigenvalues of the system are,
\begin{equation}
{E_n} = \hbar {\omega _a}{n_a} + \hbar {\omega _b}{n_b} + \frac{{\hbar \kappa }}{2}{n_a}{n_b}.\tag{4}
\end{equation}
To obtain the deformed function corresponding to this two-mode nonlinear system, we try to describe this system with the following two-dimensional f-deformed oscillator Hamiltonian:
\begin{equation}
  \hat H = \hbar {\omega _a}{\hat A^\dag }\hat A + \hbar {\omega _b}{\hat b^\dag }\hat b,\tag{5}
\end{equation}
whose eigenvalues are given by
\begin{equation}
  {E_n} = \hbar {\omega _a}f_a^2({n_a},{n_b}){n_a} + \hbar {\omega _b}{n_b}.\tag{6}
\end{equation}
If we now compare the energy spectrum of the above mentioned nonlinear medium (4) with the energy spectrum of the two-dimensional deformed oscillator (6), we obtain the corresponding deformation function as,
\begin{equation}
  {\hat f_a}(\kappa ,{\hat n_b}) = {\left( {1 + \frac{\kappa }{{2{\omega _a}}}{{\hat n}_b}} \right)^{\frac{1}{2}}}.\tag{7}
\end{equation}
As it is seen, the deformation function ${\hat f_a}$  is a function of number operator of the second mode, ${n_b}$, and also the parameter of nonlinearity $\kappa $. Finally, the deformed bosonic oscillator operators are given by,\\
\begin{equation}
 \begin{split}
 \begin{array}{l}\hat A = \hat a\sqrt {1 + \frac{\kappa }{{2{\omega _a}}}{{\hat n}_b}},\,\,\,\,\,\,\,\ {{\hat A}^\dag } = \sqrt {1 + \frac{\kappa }{{2{\omega _a}}}{{\hat n}_b}} {{\hat a}^\dag }.\end{array}
 \end{split}\tag{8}
 \end{equation}

\subsection{$\kappa $-dependent deformed $su(2)$ algebra}

As an alternative point of view, we show that our two-mode nonlinear optical system can also be considered as a new type of deformed $su(2)$ algebra. As is known, the generators of the $su(2)$ algebra satisfy the following commutation relations ~\cite{schwinger1965angular}
\begin{equation}
 \begin{split}
\begin{array}{l}\left[ {{{\hat J}_0},{{\hat J}_ \pm }} \right] =  \pm {{\hat J}_ \pm },\,\,\,\,\, \left[ {{{\hat J}_ + },{{\hat J}_ - }} \right] = 2{{\hat J}_0}.\end{array}
 \end{split}\tag{9}
 \end{equation}
Schwinger showed that the generators of the $su(2)$ algebra may be described by means of the occupation number representation of the two-dimensional isotropic harmonic oscillator. In term of the Jordan-Schwinger mapping, the generators of $su(2)$ algebra may be respectively realized by two pairs of mutually commuting boson operators $\{ \hat a,{\hat a^\dag }\} $ and $\{ \hat b,{\hat b^\dag }\} $ as
\begin{equation}
 \begin{split}
\begin{array}{lr}
{{\hat J}_ + } = {{\hat a}^\dag }\hat b,\,\,\,\ {{\hat J}_ - } = {{\hat b}^\dag }\hat a,\,\,\,\ {{\hat J}_0} = \frac{1}{2}\left( {{{\hat n}_a} - {{\hat n}_b}} \right),
\end{array}
\end{split}\tag{10}
 \end{equation}
where
\begin{equation}
 \begin{split}
\begin{array}{l}{{\hat n}_a} = {{\hat a}^\dag }\hat a,\,\,\,\,\,\ {{\hat n}_b} = {{\hat b}^\dag }\hat b\,.\end{array}
 \end{split}\tag{11}
 \end{equation}
In order to construct a deformed $su(2)$ algebra corresponding to the cross-Kerr medium, we define f-deformed (nonlinear) two-mode realization for its generators, in the similar approach of introducing the f-deformed algebra, as
\begin{equation}
 \begin{split}
\begin{array}{l}\hat J_ + ^{(\kappa )} = {{\hat A}^\dag }\hat b = {{\hat f}_a}(\kappa ,{{\hat n}_b}){{\hat a}^\dag }\hat b,\\\hat J_ - ^{(\kappa )} = {{\hat b}^\dag }\hat A = {{\hat b}^\dag }\hat a{{\hat f}_a}(\kappa ,{{\hat n}_b}),\\\hat J_0^{(\kappa )} = \frac{1}{2}\left( {{{\hat n}_a} - {{\hat n}_b}} \right).\end{array}
 \end{split}\tag{12}
 \end{equation}
It can be easily shown that the above deformed operators satisfy the following deformed $s{u_\kappa }(2)$ algebra:
\begin{equation}
 \begin{split}
\begin{array}{l}\left[ {\hat J_0^{(\kappa )},\hat J_ \pm ^{(\kappa )}} \right] =  \pm \hat J_ \pm ^{(\kappa )},\\[10pt]
\left[ {\hat J_ + ^{(\kappa )},\hat J_ - ^{(\kappa )}} \right] = 2\hat J_0^{(\kappa )} + O(\kappa ).\end{array}
 \end{split}\tag{13}
 \end{equation}
It is clear that in the limit $\kappa  \to 0, {f_a}(\kappa ,{n_b}) \to 1$ and the above $s{u_\kappa }(2)$ algebra reduces to the standard
nondeformed $su(2)$ algebra, Eq. (9).

\section{ Nonlinear coherent state for the nonlinear optical system}

By using the definition of nonlinear two-mode realization of the $s{u_\kappa }(2)$, Eq. (12), we obtain
\begin{equation}
 \begin{split}
\begin{array}{l}\hat J_ - ^{(\kappa )}\left| {0,N} \right\rangle  = {{\hat b}^\dag }\hat a\sqrt {1 + \frac{\kappa }{{2{\omega _a}}}{{\hat n}_b}} \left| {0,N} \right\rangle  = 0,\\\\\hat J_ + ^{(\kappa )}\left| {N,0} \right\rangle  = \sqrt {1 + \frac{\kappa }{{2{\omega _a}}}{{\hat n}_b}} {{\hat a}^\dag }\hat b\left| {N,0} \right\rangle  = 0.\end{array}
 \end{split}\tag{14}
 \end{equation}
The first equation states that  $\left| {0,N} \right\rangle  \equiv {\left| 0 \right\rangle _a} \otimes {\left| N \right\rangle _b}$ is the lowest weight state of the $s{u_\kappa }(2)$ algebra. It is also obvious from the second equation that, for each constant value of $N$, we encounter with a finite-dimensional Hilbert space. In this section, we are intended to construct CSs associated with our nonlinear optical system. We can make use of the formalism of truncated CSs~\cite{kuang1993dynamics} and define cross-Kerr nonlinear CSs (CK-NCSs) corresponding to the nonlinear two-mode realization of the $s{u_\kappa }(2)$ algebra as
\begin{equation}
\begin{split}
\begin{array}{ll}{\left| \mu  \right\rangle _\kappa } = {C^{ - 1}}\exp (\mu \hat J_ + ^{(\kappa )})\left| {0,N} \right\rangle \\[10pt]
\,\,\,\, = {C^{ - 1}}\exp (\mu \sqrt {1 + \frac{\kappa }{{2{\omega _a}}}{{\hat n}_b}} {{\hat a}^\dag }\hat b)\left| {0,N} \right\rangle \\[10pt]
\,\,\,\, = \,{C^{ - 1}}\sum\limits_{n = 0}^N {\sqrt {\left( {\begin{array}{*{20}{c}}N\\n\end{array}} \right)} } {f_a}(\kappa ,{n_b})!\,{\mu ^n}\left| {n,N - n} \right\rangle ,\end{array}
\end{split}\tag{15}
\end{equation}
where $\mu $ is a complex number and
\begin{equation}
{C^2} = \sum\limits_{n = 0}^N {\left( {\begin{array}{*{20}{c}}N\\n\end{array}} \right)} {\left[ {{f_a}(\kappa ,{n_b})!} \right]^2}\,{\left( {{{\left| \mu  \right|}^2}} \right)^n}.\tag{16}
\end{equation}
By definition
\begin{equation}
\begin{array}{l}\left[ {{f_a}(\kappa ,0)} \right]! = 1,\\[10pt]
\left[ {{f_a}(\kappa ,{n_b})} \right]! = {f_a}(\kappa ,{n_b})\left[ {{f_a}(\kappa ,{n_b} - 1)} \right]!,\end{array}\tag{17}
\end{equation}
and ${f_a}(\kappa ,{n_b})$ is given by Eq. (7).
It is clear that in the limit $\kappa  \to 0, {f_a}(\kappa ,{n_b}) \to 1$ and above deformed CK-NCSs reduce to the CSs for bosonic realization of the $su(2)$ algebra~\cite{buvzek1989generalized}.
It is worth noting that the two-mode states $\left| {n,N - n} \right\rangle  = {\left| n \right\rangle _a} \otimes {\left| {N - n} \right\rangle _b}$ have the following properties
\begin{equation}
\begin{array}{l}{{\hat n}_a}\left| {n,N - n} \right\rangle  = n\left| {n,N - n} \right\rangle ,\\{{\hat n}_b}\left| {n,N - n} \right\rangle  = (N - n)\left| {n,N - n} \right\rangle ,\\\left\langle {n,N - n\left| {m,N - m} \right.} \right\rangle  = {\delta _{nm}},\\\sum\limits_n^N {\left| {n,N - n} \right\rangle \left\langle {n,N - n} \right|}  = 1.\end{array}\tag{18}
\end{equation}
Therefore, two-mode nonlinear CSs, Eq. (15), can be considered as a two-mode entangled state. If $n$ photons are in one mode, $N - n$ photons are located in the other mode, i.e. the total number of the photon equaled to $N$.

\subsection{Resolution of identity}

In this section, we show that the CK-NCSs form an overcomplete set. Since it is necessary to include a measure function $m({\left| \mu  \right|^2})$in the integral, we require
\begin{equation}
\int {{d^2}\mu \left| \mu  \right\rangle } m({\left| \mu  \right|^2})\left\langle \mu  \right| = \sum\limits_{n = 0}^N {\left| n, N-n \right\rangle \left\langle n, N-n \right|}  = \hat {\rm I}. \tag{19}
\end{equation}
In the case of $\kappa  = 0$,
\begin{equation}
\begin{split}
\begin{array}{l}{\int {{d^2}\mu \left| \mu  \right\rangle } _{\kappa  = 0}}{m_{\kappa  = 0}}({\left| \mu  \right|^2}){}_{\kappa  = 0}\left\langle \mu  \right| \\[10pt]
=\sum\limits_{n = 0}^N {\left( {\begin{array}{*{20}{c}}N\\n\end{array}} \right)} \left| {n, N-n} \right\rangle \left\langle {n, N-n} \right| \\[10pt]
\times{\int {d(\left| \mu  \right|} ^2})d\theta \frac{{{{\left| \mu  \right|}^2}}}{{{{(1 + {{\left| \mu  \right|}^2})}^N}}}{m_{\kappa  = 0}}({\left| \mu  \right|^2})\\[10pt]
=\pi \sum\limits_{n = 0}^N {\left( {\begin{array}{*{20}{c}}N\\n\end{array}} \right)} \left| {n, N-n} \right\rangle \left\langle {n, N-n} \right|\\[10pt]
\times{\int_0^\infty  {d(\left| \mu  \right|} ^2})d\theta \frac{{{{\left| \mu  \right|}^2}}}{{{{(1 + {{\left| \mu  \right|}^2})}^N}}}{m_{\kappa  = 0}}({\left| \mu  \right|^2}).\end{array}
\end{split}\tag{20}
\end{equation}
Thus, we should have
\begin{equation}
\begin{split}\begin{array}{l}
{\int_0^\infty  {d(\left| \mu  \right|} ^2})d\theta \frac{{{{\left| \mu  \right|}^2}}}{{{{(1 + {{\left| \mu  \right|}^2})}^N}}}{m_{\kappa  = 0}}({\left| \mu  \right|^2}) = \frac{1}{{\pi \left( {\begin{array}{*{20}{c}}N\\n\end{array}} \right)}}.\end{array}
\end{split}\tag{21}
\end{equation}
The suitable choice for the measure function reads~\cite{mahdifar2006geometric}
\begin{equation}
{m_{\kappa  = 0}}({\left| \mu  \right|^2}) = \frac{{N + 1}}{\pi }\frac{{{{\left| \mu  \right|}^2}}}{{{{(1 + {{\left| \mu  \right|}^2})}^2}}},\tag{22}
\end{equation}
so, the resolution of identify is
\begin{equation}
\frac{{N + 1}}{\pi }{\int {d(\left| \mu  \right|} ^2}){\left| \mu  \right\rangle _{\kappa  = 0}}{}_{\kappa  = 0}\left\langle \mu  \right| = \hat I.\tag{23}
\end{equation}
In order to examine the resolution of identity for the CK-NCSs, we first define the deformed Binomial expansion ~\cite{mahdifar2006geometric}
\begin{equation}
\left( {1 + x} \right)_\kappa ^N = {\sum\limits_{n = 0}^N {\left( {\begin{array}{*{20}{c}}N\\n\end{array}} \right)} _\kappa }\,{x^n},\tag{24}
\end{equation}
where
\begin{equation}
{\left( {\begin{array}{*{20}{c}}N\\n\end{array}} \right)_\kappa } = \left( {\begin{array}{*{20}{c}}N\\n\end{array}} \right){\left\{ {\left. {\left| {{f_a}(\kappa ,{n_b})!} \right|} \right\}} \right.^2}.\tag{25}
\end{equation}
We see that when $\kappa  \to 0,\,\,\,{f_a}(\kappa ,{n_b}) \to 1$ and deformed Binomial expansion becomes the well-known Binimial expansion. Now, using this definition we can rewrite the CK-NCSs as,
\begin{equation}
{\left| \mu  \right\rangle _\kappa } =
\left( {1 + {{\left| \mu  \right|}^2}} \right)_\kappa ^{ - {N \mathord{\left/ {\vphantom {N 2}} \right.
 \kern-\nulldelimiterspace} 2}}\sum\limits_{n = 0}^N {\sqrt {{{\left( {\begin{array}{*{20}{c}}N\\n\end{array}} \right)}_\kappa }} } {\mu ^n}\left| {n,N - n} \right\rangle . \tag{26}
\end{equation}
For the resolution of identify we should have
\begin{equation}
\begin{split}
\begin{array}{l}
\int_{(\kappa )} {{d^2}\mu {{\left| \mu  \right\rangle }_\kappa }} {m_\kappa}({\left| \mu  \right|^2}){}_\kappa \left\langle \mu \right|=\sum\limits_{n = 0}^N {\left| {n,N - n} \right\rangle \left\langle {n,N - n} \right| = } \hat {\rm I},\end{array}
\end{split}\tag{27}
\end{equation}
or, equivalently
\begin{equation}
\begin{split}\begin{array}{l}
\pi \sum\limits_{n = 0}^N {{{\left( {\begin{array}{*{20}{c}}N\\n\end{array}} \right)}_\kappa }} \left| {n, N-n} \right\rangle \left\langle {n, N-n} \right|\\[10pt]
\times{\int_0^\infty  {d(\left| \mu  \right|} ^2})d\theta \frac{{{{\left| \mu  \right|}^2}}}{{(1 + {{\left| \mu  \right|}^2})_\kappa ^N}}{m_\kappa }({\left| \mu  \right|^2}) = \hat I.\end{array}
\end{split}\tag{28}
\end{equation}
If we define the following measure,
\begin{equation}
{m_\kappa }({\left| \mu  \right|^2}) = \frac{{N + 1}}{\pi }\frac{1}{{\left( {1 + {{\left| \mu  \right|}^2}} \right)_\kappa ^2}},\tag{29}
\end{equation}
and the following deformed version of Eq. (23)
\begin{equation}
{\int_{0(\kappa )}^\infty  {d(\left| \mu  \right|} ^2})\frac{{{{\left| \mu  \right|}^{2n}}}}{{(1 + {{\left| \mu  \right|}^2})_\kappa ^2}}\left| \mu  \right\rangle {}_\kappa {}_\kappa \left\langle \mu  \right| = \frac{1}{{\pi {{\left( {\begin{array}{*{20}{c}}N\\n\end{array}} \right)}_\kappa }}},\tag{30}
\end{equation}
we arrive at the resolution of identify for the ${\left| \mu  \right\rangle _\kappa }$:
\begin{equation}
\frac{{N + 1}}{\pi }\int_{(\kappa )} {\frac{{{d^2}\mu }}{{\left( {1 + {{\left| \mu  \right|}^2}} \right)_\kappa ^2}}} {\left| \mu  \right\rangle _\kappa }{}_\kappa \left\langle \mu  \right| = \hat {\rm I}.\tag{31}
\end{equation}

\section{Quantum statistical properties of the CK-NCSs}

In the present section, we study some quantum statistical properties of the two-mode nonlinear CSs for our nonlinear optical system, including mean number of photons, cross-correlation and Mandel parameter.

\subsection{photon number distribution}

The mean number of photons in two modes are given by
\begin{equation}
\begin{array}{l}\left\langle {{{\hat n}_a}} \right\rangle = Tr({{\hat \rho }_a}\hat n) \\[10pt]
= {C^{ - 2}}\sum\limits_{n = 0}^N {\left( {\begin{array}{*{20}{c}}N\\n\end{array}} \right)} {\left[ {f(\kappa ,n)!} \right]^2}\,{\left( {{{\left| \mu  \right|}^2}} \right)^n}n,\\[10pt]
\\\left\langle {{{\hat n}_b}} \right\rangle  = Tr({{\hat \rho }_b}\hat n) \\[10pt]
= {C^{ - 2}}\sum\limits_{n = 0}^N {\left( {\begin{array}{*{20}{c}}N\\n\end{array}} \right)} {\left[ {f(\kappa ,n)!} \right]^2}\,{\left( {{{\left| \mu  \right|}^2}} \right)^n}(N - n).\end{array}\tag{32}
\end{equation}

In Figs. 1(a) and 1(b), respectively, we have plotted the variation the mean numbers of photons in the first and second modes of the state ${\left| \mu  \right\rangle _\kappa }$ with respect to the dimensionless parameter $\tilde \kappa  = \frac{\kappa }{{2{\omega _a}}}$ for different values of $\mu $. We observe that for a given $\mu $, the mean number of photons in the first (second) mode is increased (decreased) by increasing $\tilde \kappa $. Also for a fixed $\tilde \kappa $, by increasing $\mu $, the mean number of photons in first (second) mode is increased (decreased).
\begin{figure}[htb]
  \centering
   \begin{subfigure}[b]{0.49\linewidth}
  \includegraphics[width=\linewidth]{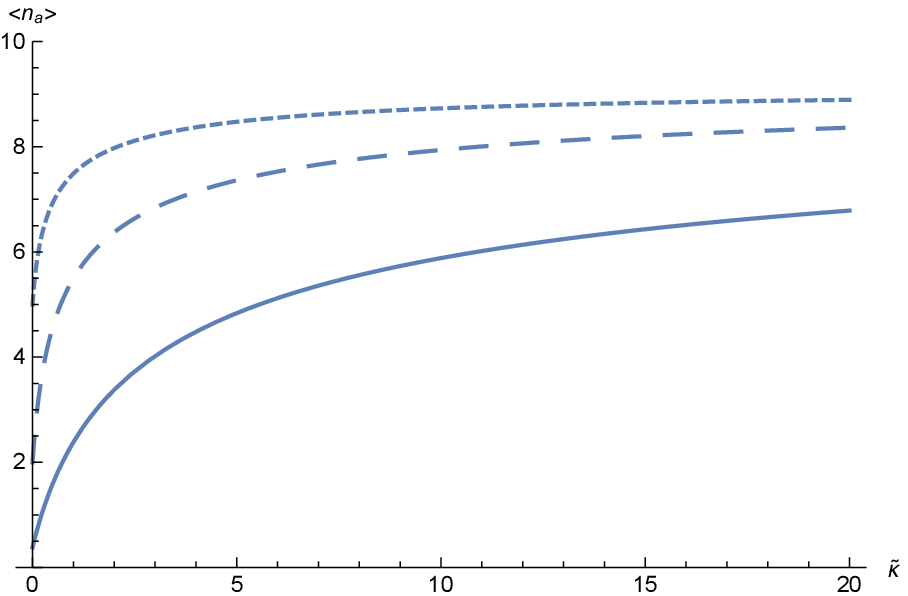}
  \caption{}
  \end{subfigure}
   \begin{subfigure}[b]{0.49\linewidth}
  \includegraphics[width=\linewidth]{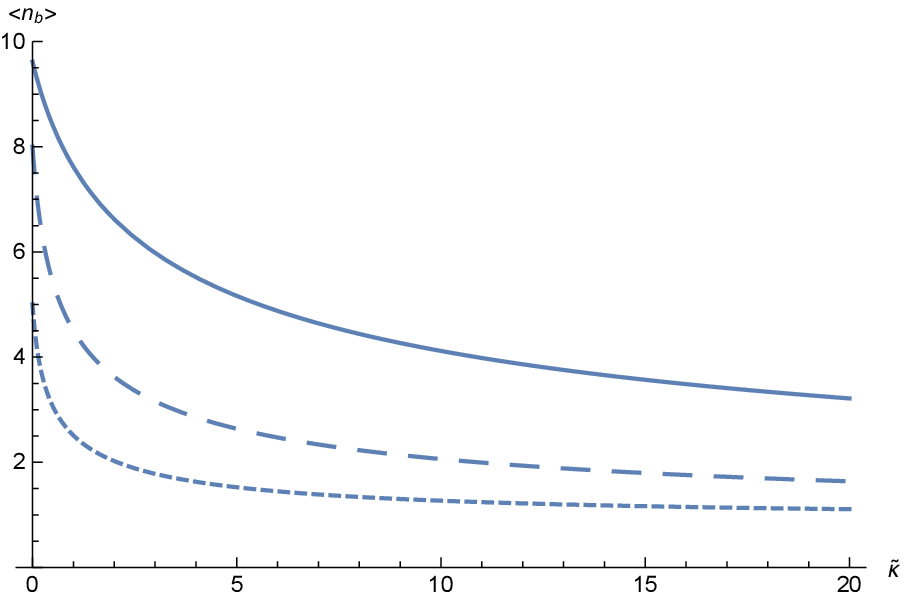}
  \caption{}
  \end{subfigure}
  \captionsetup{width=1\linewidth, font=footnotesize}
  \caption{Mean number of photons for (a) the first mode and (b) the second mode of the state $|\mu\rangle_\kappa$ versus $\tilde \kappa$ for $\mu= 0.2$ (solid line), $\mu= 0.5$ (dashed line) and $\mu= 1$ (dotted line), with $N=10$.}

\end{figure}

One of the interesting properties of the two-mode CK-NCSs is the anticorrelation between the modes. This anticorrelation is defined by the normalized cross-correlation function as~\cite{buvzek1989generalized}
\begin{equation}
{g^2} = \frac{{\left\langle {{n_a}{n_b}} \right\rangle }}{{\left\langle {{n_a}} \right\rangle \left\langle {{n_b}} \right\rangle }}.\tag{33}
\end{equation}
We use the following equation for CK-NCSs
\begin{equation}
\begin{split}\begin{array}{l}
\left\langle {{{\hat n}_a}{{\hat n}_b}} \right\rangle={C^{ - 2}}\sum\limits_{n = 0}^N {\left( {\begin{array}{*{20}{c}}N\\n\end{array}} \right)} {\left[ {f(\kappa ,n)!} \right]^2}\,{\left( {{{\left| \mu  \right|}^2}} \right)^n}n(N - n),\end{array}
\end{split}\tag{34}
\end{equation}
and calculate the cross-correlation function. In Fig. 2, we have plotted the cross-correlation function with respect to $N$, for $\tilde \kappa  = 0.1$ and $\mu  = 0.1.$
\begin{figure}[h]
 \centering
  \includegraphics[width=1\linewidth]{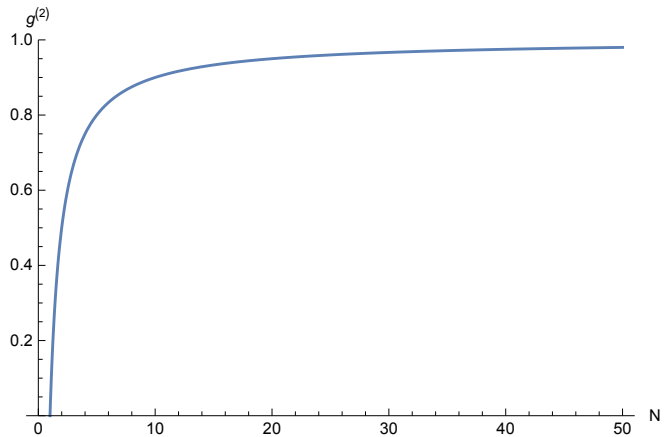}
    \captionsetup{width=1\linewidth, font=footnotesize}
  \caption{Cross-correlation function for the state $|\mu\rangle_\kappa$ with respect $N$ for $\tilde \kappa =0.1$ and $\mu=0.1$.}

\end{figure}
In this case, for any values of $N$, the cross-correlation function is less than one. This indicates that the two modes which are discussed are anticorrelated. Physically this means that there is no tendency for bosons in the different modes to be created or annihilated simultaneously. Furthermore, in the limit of $N \to \infty ,\,\,{g^{(2)}}$ is equal to one, which indicates that the bosons in the different modes are created or annihilated independtly.
\subsection{ Sub-Poissonian statistics }
In order to determine the quantum statistics of the CK-NCSs, we consider Mandel parameter for two modes of ${\left| \mu  \right\rangle _\kappa }$ as~\cite{mandel1995optical}\\
\begin{equation}
{Q_i} = \frac{{{{\left( {\Delta {n_i}} \right)}^2} - \left\langle {{{\hat n}_i}} \right\rangle }}{{\left\langle {{{\hat n}_i}} \right\rangle }},\,\,\,\,\,\,\,\,i = a,b.\tag{35}
\end{equation}
Negative values of $Q$ correspond to states which their variance of photon is less than mean (sub-Poissonian and photon antibunching). Also, $Q$ is positive for a super-Poissonian distribution (photon bunching) and $Q = 0$ corresponds to Poissonian distribution. Figs. 3(a) and 3(b) show the effect of nonlinearity on the variation of Mandel parameters of two modes of our nonlinear CSs with $N = 10$ for different values of $\mu $.\\
\begin{figure}[htb]
  \centering
  \begin{subfigure}[b]{0.49\linewidth}
    \includegraphics[width=\linewidth]{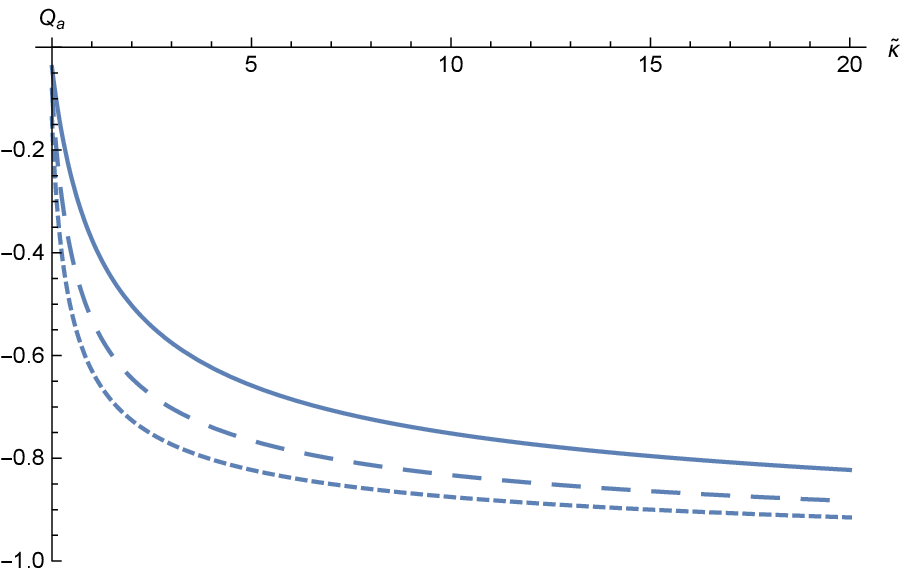}
    \caption{}
  \end{subfigure}
  \begin{subfigure}[b]{0.49\linewidth}
    \includegraphics[width=\linewidth]{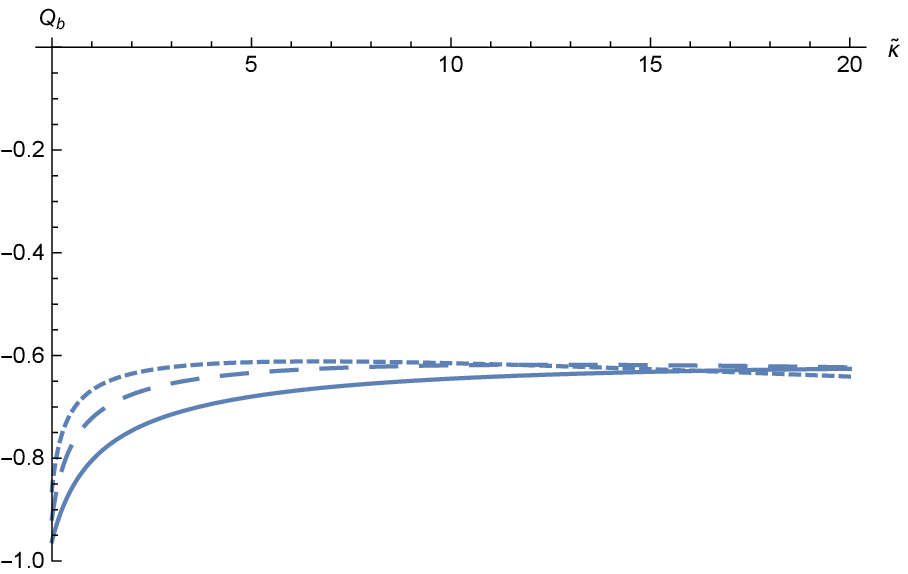}
    \caption{}
  \end{subfigure}
    \captionsetup{width=1\linewidth, font=footnotesize}
  \caption{Mandel parameter of (a) the first mode and (b) the second mode of the state $|\mu\rangle_\kappa$ versus $\tilde \kappa$ for $\mu= 0.2$ (solid line), $\mu= 0.3$ (dashed line) and $\mu= 0.4$ (dotted line), with $N=10$.}

\end{figure}

As it is seen, both modes have the sub-Poissonian statistics, although by increasing $\tilde \kappa $, photon-counting statistics of the first mode of the CSs tends to sub-Poissonian more rapidly. In other words, the first mode of the CSs, related to a nonlinear media shows more nonclassical properties than the CSs of a linear media.

On the other hand, by increasing $\tilde \kappa $, the Mandel parameter of the second mode increases to a maximum and then decreases to -1. In other words, with increasing nonlinearity of the media, the statistics of the second mode of the constructed CSs, tends to Poissonian for small values of nonlinear parameter $\tilde \kappa $, and tends to sub-Poissonian for large values of $\tilde \kappa $. In summary, two modes of the CSs of the nonlinear media with large nonlinearity, show more nonclassical properties than the modes of the CSs of the media with small nonlinearity.

\section{Interaction Hamiltonian and the time-dependent atom-field state}

In this section, as an example of applicability of the constructed CK-NCSs, we try to analyze the nonlinear effects of the medium on the dynamics of atom-field interaction within the framework of the NCSs. For this purpose, first, we should prepare a three-level  atom in a cross-Kerr nonlinear media. The cross-Kerr nonlinearity could be obtained by the interaction of the signal and the probe fields in a four-level N-type atoms system~\cite{hu2011cross} which its Hamiltonian is given by Eq. (3). Then, we drive the three-level atom with these two-mode fields. As mentioned before, we will consider the nonlinear effects of the nonlinear media via the indirect CK-NCSs approach and extract some physical features that are originated from the nonlinearity of the media. We will find the effects of nonlinearity on the atomic occupation probabilities, the cross-correlation function, the Mandel parameter, the squeezing and the von Neumann entropy.

To this end, we consider the three-level atom in a $\Lambda $ configuration as shown in Fig. 4. The possible transitions are $0 \leftrightarrow 1$ and $1 \leftrightarrow 2$. The interaction Hamiltonian of the system can be expressed in the rotating-wave approximation and in the interaction picture as follows~\cite{ashby2003relativity}:
\begin{equation}
{\hat H_I} = {g_a}\left( {{{\hat \sigma }_{10}}\hat a + {{\hat a}^\dag }{{\hat \sigma }_{01}}} \right) + {g_b}\left( {{{\hat \sigma }_{12}}\hat b + {{\hat b}^\dag }{{\hat \sigma }_{21}}} \right).\tag{36}
\end{equation}
Here $\hat a({\hat a^\dag })$ and $\hat b({\hat b^\dag })$ are the annihilation (creation) operators of the field modes, respectively, ${\hat \sigma _{ij}} = \left| i \right\rangle \left\langle j \right|$ are the atomic raising and lowering operators, ${g_a}$ and ${g_b}$ are the atom-field coupling strengths for each transition channel.
 \begin{figure}[h]
 \centering
   \includegraphics[width=0.6\linewidth]{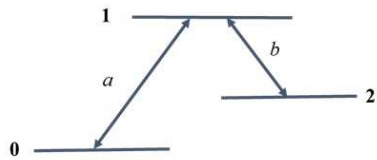}
     \captionsetup{width=1\linewidth, font=footnotesize}
  \caption{Energy diagram of a three-level atom in the $\Lambda$ configuration interaction with two quantized cavity modes.}
\end{figure}
To obtain the nonlinear effects of the media on the dynamics of this system and describe the time-dependent state of the system, we suppose the atom is initially prepared in the ground state $\left| 0 \right\rangle $, and the field is initially prepared in the two-mode CK-NCSs of the nonlinear media, Eq. (15), as:
\begin{equation}
{\left| {\psi (0)} \right\rangle _f} = \sum\limits_{{n_a},{n_b}}^N {{A_{{n_a},{n_b}}}} (\kappa )\left| {{n_a},{n_b}} \right\rangle ,\tag{37}
\end{equation}
where ${\left| {{A_{{n_a},{n_b}}}(\kappa )} \right|^2}$ is the  joint probability of finding ${n_a}$ photon in the mode $a$ and ${n_b}$ photon in the mode $b$ of two-mode CK-NCSs:
\begin{equation}
\begin{split}
{\left| {{A_{{n_a},{n_b}}}(\kappa )} \right|^2} = {p_f}({n_a},{n_b};\kappa ) = {\left| {\left\langle {{n_a},{n_b}} \right|{{\left. \mu  \right\rangle }_\kappa }} \right|^2} \\[10pt]
= {C^{ - 2}}\left( {\begin{array}{*{20}{c}}N\\{{n_a}}\end{array}} \right){\left[ {{f_a}(\kappa ,{n_b})!} \right]^2}\,{\left( {{{\left| \mu  \right|}^2}} \right)^n}{\delta _{{n_b},N - {n_a}}}.
\end{split}\tag{38}
\end{equation}
We write the atom-field state of the system at any time $t > 0$ as follows:
\begin{equation}
\begin{array}{l}\left| {\psi (t)} \right\rangle  = \sum\limits_{{n_a},{n_b}} {{A_{{n_a},{n_b}}}} (\kappa )\\
 \,\,\,\,\,\,\,\,\,\,\,\,\,\,\,\,\,\,\,\,\,\,\,\times\{ {C_0}({n_a},{n_b};t)\left| {0;{n_a},{n_b}} \right\rangle \\
  \,\,\,\,\,\,\,\,\,\,\,\,\,\,\,\,\,\,\,\,\,\,\,+{C_1}({n_a},{n_b};t)\left| {1;{n_a} - 1,{n_b}} \right\rangle\\ \,\,\,\,\,\,\,\,\,\,\,\,\,\,\,\,\,\,\,\,\,\,\,+{C_2}({n_a},{n_b};t)\left| {2;{n_a} - 1,{n_b} + 1} \right\rangle \} ,\,\,\,\,\,\,\,\,\,\,\end{array}\tag{39}
\end{equation}
where ${C_i}({n_a},{n_b};t)$ are the atomic probability amplitudes, and by using the time-dependent Schrödinger equation in the interaction picture,
\begin{equation}
i\hbar \frac{\partial }{{\partial t}}\left| {\psi (t)} \right\rangle  = {\hat H_I}\left| {\psi (t)} \right\rangle ,\tag{40}
\end{equation}
are obtained from:
\begin{equation}
\begin{array}{l}i{{\dot C}_0} = {\Omega _{{n_a} - 1}}{C_1},\,\\i{{\dot C}_1} = {\Omega _{{n_a} - 1}}{C_0} + {\Omega _{{n_b}}}{C_2},\,\\i{{\dot C}_2} = {\Omega _{{n_b}}}{C_1},\end{array}\tag{41}
\end{equation}
where ${\Omega _{{n_a}}}$ and ${\Omega _{{n_b}}}$ are the one-photon Rabi frequencies defined by
\begin{equation}
{\Omega _{{n_a}}} = {g_a}{\left[ {{n_a} + 1} \right]^{\frac{1}{2}}},\,\,\,\,\,\,\,{\Omega _{{n_b}}} = {g_b}{\left[ {{n_b} + 1} \right]^{\frac{1}{2}}}.\tag{42}
\end{equation}
By calculating the solutions of Eqs. (41), we can obtain the occupation probabilities of the various levels in the two modes $\left( {{P_i}({n_a},{n_b};t) = {{\left| {{C_i}({n_a},{n_b};t)} \right|}^2}} \right)$ as bellow:
\begin{equation}
\begin{array}{l}{P_0}({n_a},{n_b};t) = \frac{{\Omega _{{n_b}}^4}}{{\Omega _{{n_a} - 1,{n_b}}^4}} + \frac{{2\Omega _{{n_a} - 1}^2\Omega _{{n_b}}^2}}{{\Omega _{{n_a} - 1,{n_b}}^4}}\cos ({\Omega _{{n_a} - 1,{n_b}}}t)\\\,\,\,\,\,\,\,\,\,\,\,\,\,\,\,\,\,\,\,\,\,\,\,\,\,\,\,\,\,\, + \frac{{\Omega _{{n_a} - 1}^4}}{{\Omega _{{n_a} - 1,{n_b}}^4}}{\cos ^2}({\Omega _{{n_a} - 1,{n_b}}}t),\\{P_1}({n_a},{n_b};t) = \frac{{\Omega _{{n_a} - 1}^2}}{{\Omega _{{n_a} - 1,{n_b}}^2}}{\sin ^2}({\Omega _{{n_a} - 1,{n_b}}}t),\\{P_2}({n_a},{n_b};t) = \frac{{\Omega _{{n_a} - 1}^2\Omega _{{n_b}}^2}}{{\Omega _{{n_a} - 1,{n_b}}^4}}[1 - 2\cos ({\Omega _{{n_a} - 1,{n_b}}}t)\\\,\,\,\,\,\,\,\,\,\,\,\,\,\,\,\,\,\,\,\,\,\,\,\,\,\,\,\, + {\cos ^2}({\Omega _{{n_a} - 1,{n_b}}}t)],\end{array}\tag{43}
\end{equation}
where ${\Omega _{{n_a},{n_b}}}$ is the two-photon Rabi frequency defined by
\begin{equation}
{\Omega _{{n_a},{n_b}}} = {\left[ {\Omega _{{n_a}}^2 + \Omega _{{n_b}}^2} \right]^{\frac{1}{2}}} = {\left[ {{n_a} + 1 + \frac{{g_b^2}}{{g_a^2}}({n_b} + 1)} \right]^{\frac{1}{2}}}{g_a}.\tag{44}
\end{equation}
Here, $\frac{1}{2}{\Omega _{{n_a} - 1,{n_b}}}$ and ${\Omega _{{n_a} - 1,{n_b}}}$ are two Rabi frequencies, which rule the atomic dynamics and are ascribe to the one-photon and two-photon processes, respectively. As may be noted, the oscillations of the occupation probabilities have two types of the revivals due to these Rabi frequencies.

\section{Nonlinear effects on the dynamic properties of the atom}

To study the photon statistical properties of the exciting field, we can obtain the occupation probabilities of the different levels. In our atom-field system, the atomic occupation probabilities of the levels are given by
\begin{equation}
{\tilde P_i}(t,\kappa ) = \sum\limits_{{n_a},{n_b}} {{p_f}({n_a},{n_b};\kappa ){P_i}(} {n_a},{n_b};t),\,\,i = 0,1,2\tag{45}
\end{equation}
where ${p_f}({n_a},{n_b};\kappa )$ is the joint distribution of the field at $t = 0$, Eq. (38), and ${P_i}({n_a},{n_b};t)$s are given by Eq. (43).
In Fig. 5, we show the atomic occupation probabilities of the various atomic levels with respect to the ${g_a}t$ (dimensionless quantity), for different values of the nonlinearity whereas the atomic-field coupling strengths are selected equal
${{{g_b}} \mathord{\left/
 {\vphantom {{{g_b}} {{g_a} = 1}}} \right.
 \kern-\nulldelimiterspace} {{g_a} = 1}}$.
It is clear that the atomic occupation probabilities have a periodic behavior which leads by two-photon Rabi frequency (Eq. (44)).
As shown in Fig. 1, the mean photons number in the first mode $a$ increased by increasing $\tilde \kappa $ and now we show that one-photon transition will increase by increasing the $\tilde \kappa $ in Fig. 5. So we can trace this increase back to the results of the Fig. 1(a).
\begin{figure}[htb]
  \centering
  \begin{subfigure}[b]{0.49\linewidth}
    \includegraphics[width=\linewidth]{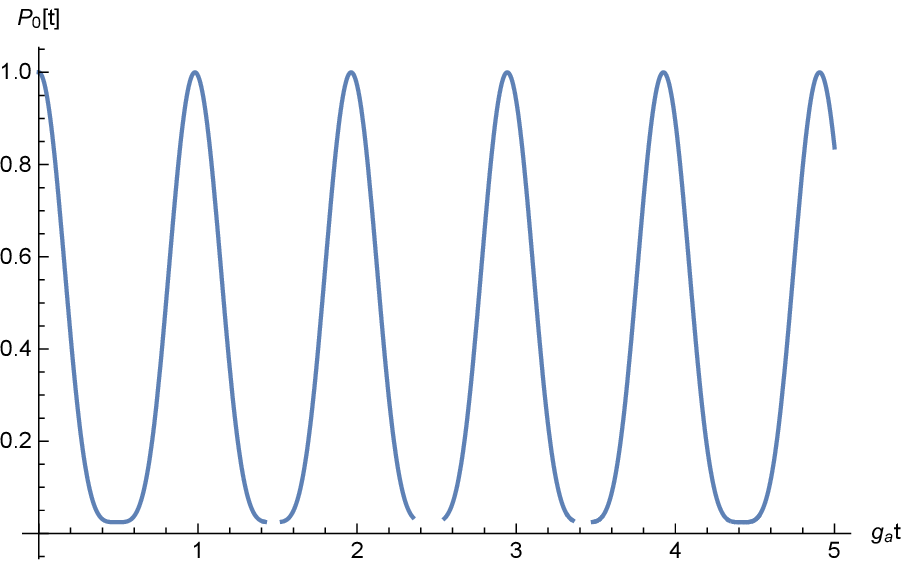}

  \end{subfigure}
  \begin{subfigure}[b]{0.49\linewidth}
    \includegraphics[width=\linewidth]{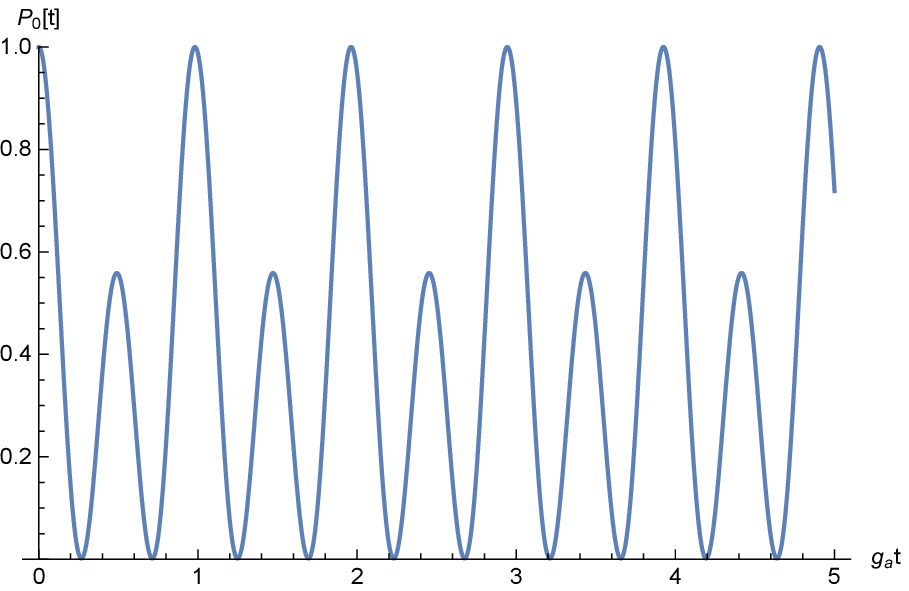}
  \end{subfigure}
  \begin{subfigure}[b]{0.49\linewidth}
    \includegraphics[width=\linewidth]{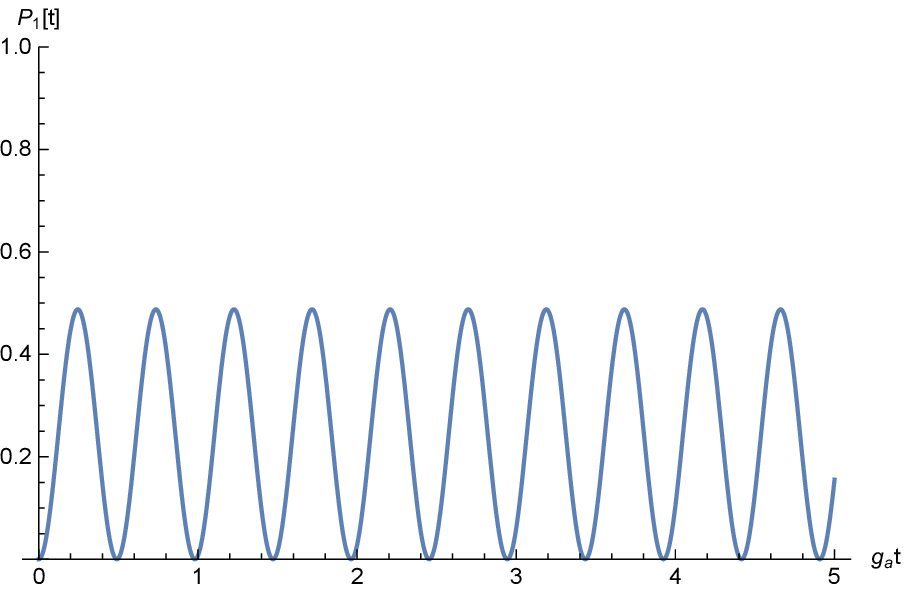}
  \end{subfigure}
  \begin{subfigure}[b]{0.49\linewidth}
    \includegraphics[width=\linewidth]{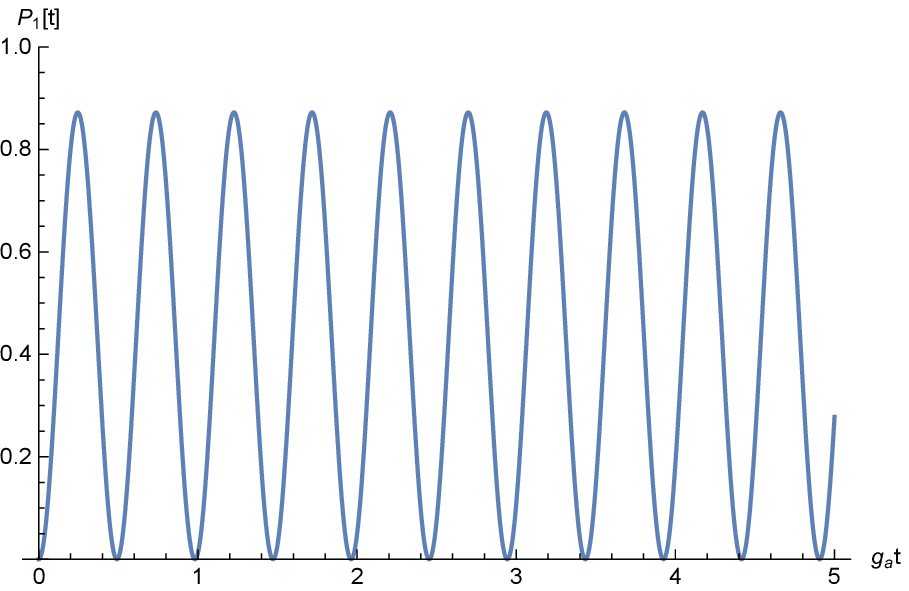}
  \end{subfigure}
    \begin{subfigure}[b]{0.49\linewidth}
    \includegraphics[width=\linewidth]{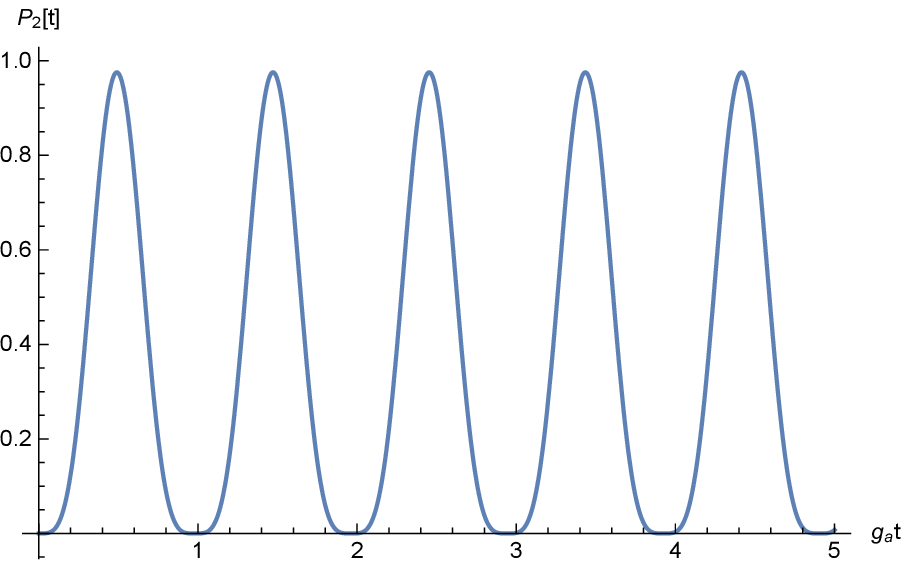}
    \caption{}
  \end{subfigure}
  \begin{subfigure}[b]{0.49\linewidth}
    \includegraphics[width=\linewidth]{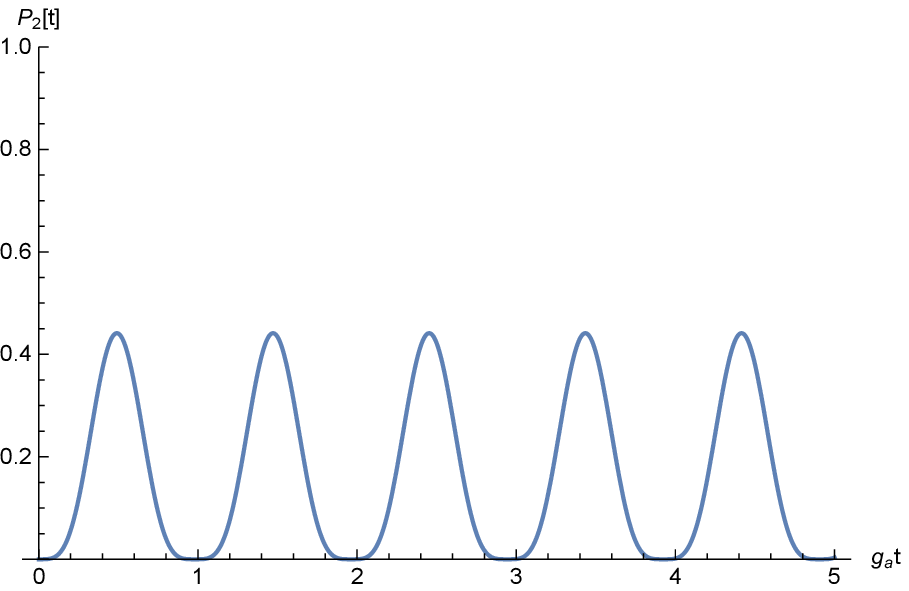}
    \caption{}
  \end{subfigure}
    \captionsetup{width=1\linewidth, font=footnotesize}
  \caption{Time evolution of the atomic occupation probabilities when the initial field is in the two-mode nonlinear CSs with $N=40$, $\mu=1$ and $g_{b}/g_{a}=1$; (a) $\tilde \kappa =0$ and (b) $\tilde \kappa =0.1$}
\end{figure}

In Fig. 6, we have plotted the atomic occupation probabilities of the various atomic levels with respect to the ${g_a}t$, for different values of the nonlinearity with nonequal atomic-field coupling strengths
${{{g_b}} \mathord{\left/
 {\vphantom {{{g_b}} {{g_a} = 2}}} \right.
 \kern-\nulldelimiterspace} {{g_a} = 2}}$.

 \begin{figure}[htb]
  \centering
  \begin{subfigure}[b]{0.49\linewidth}
    \includegraphics[width=\linewidth]{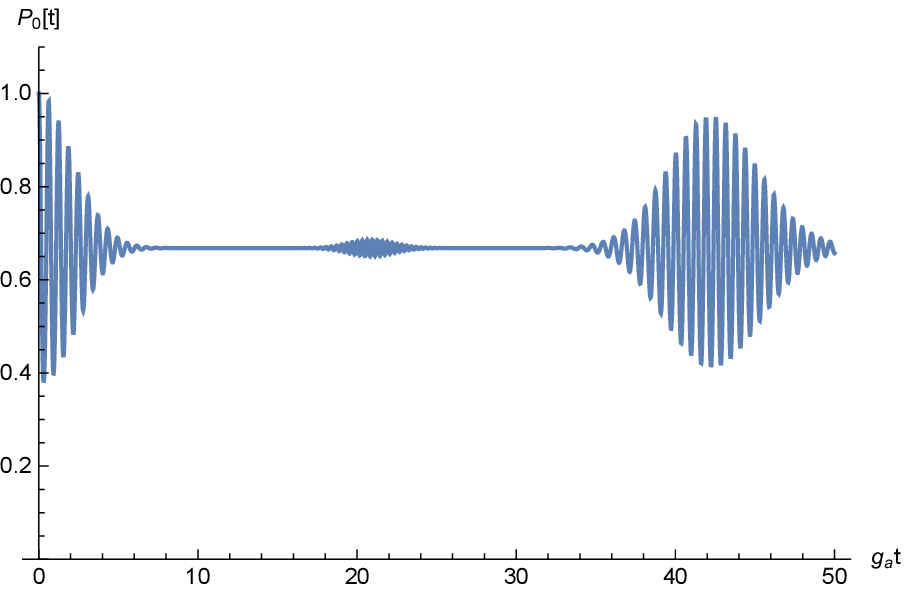}
  \end{subfigure}
  \begin{subfigure}[b]{0.49\linewidth}
    \includegraphics[width=\linewidth]{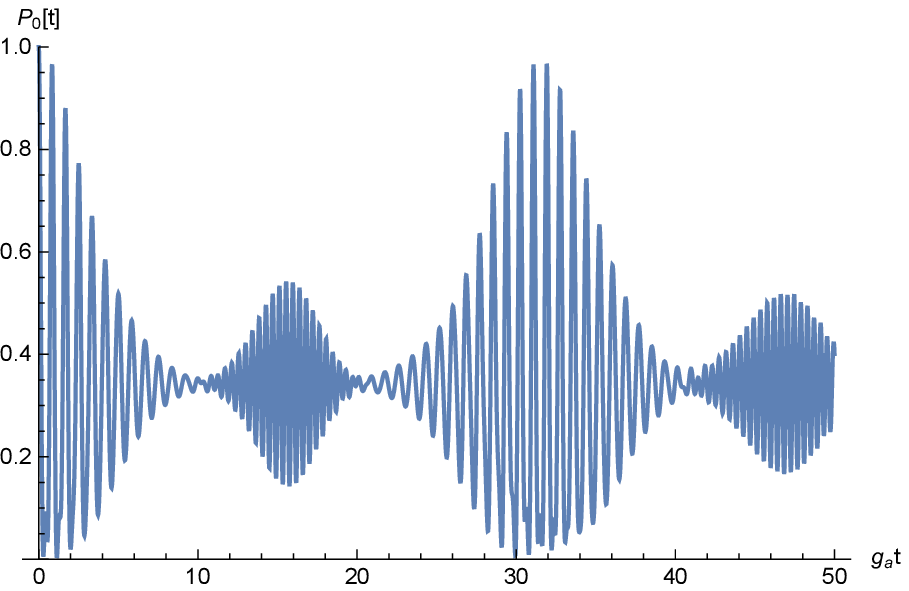}
  \end{subfigure}
  \begin{subfigure}[b]{0.49\linewidth}
    \includegraphics[width=\linewidth]{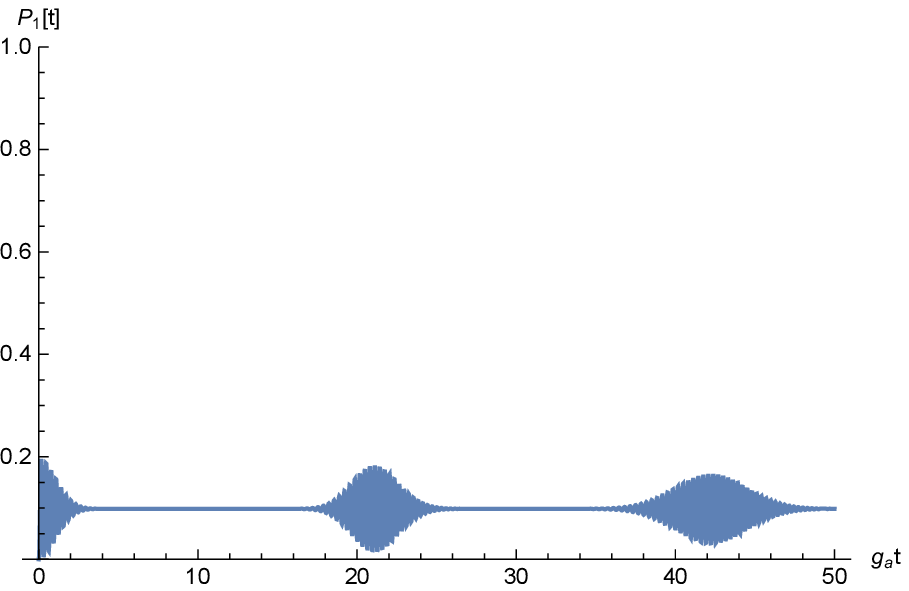}
  \end{subfigure}
  \begin{subfigure}[b]{0.49\linewidth}
    \includegraphics[width=\linewidth]{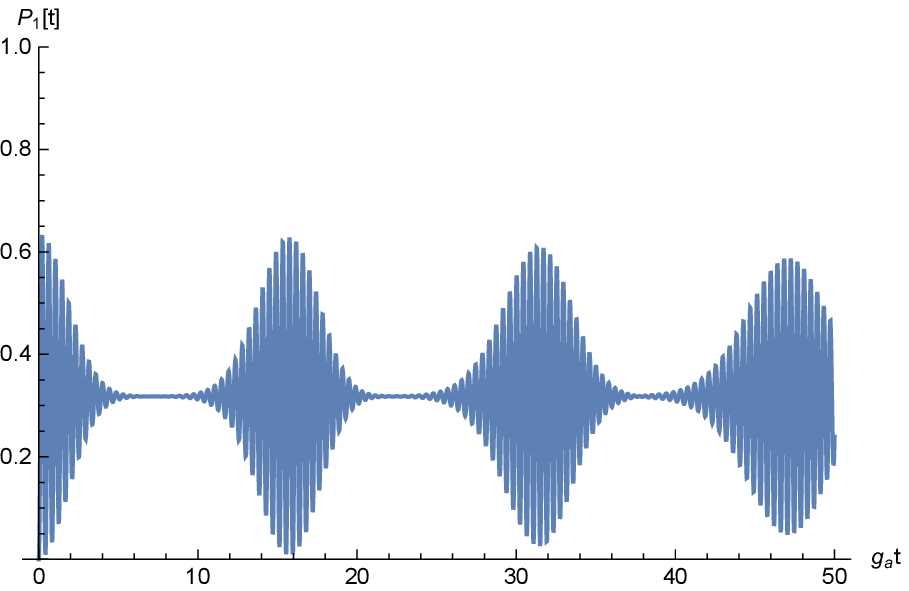}
  \end{subfigure}
    \begin{subfigure}[b]{0.49\linewidth}
    \includegraphics[width=\linewidth]{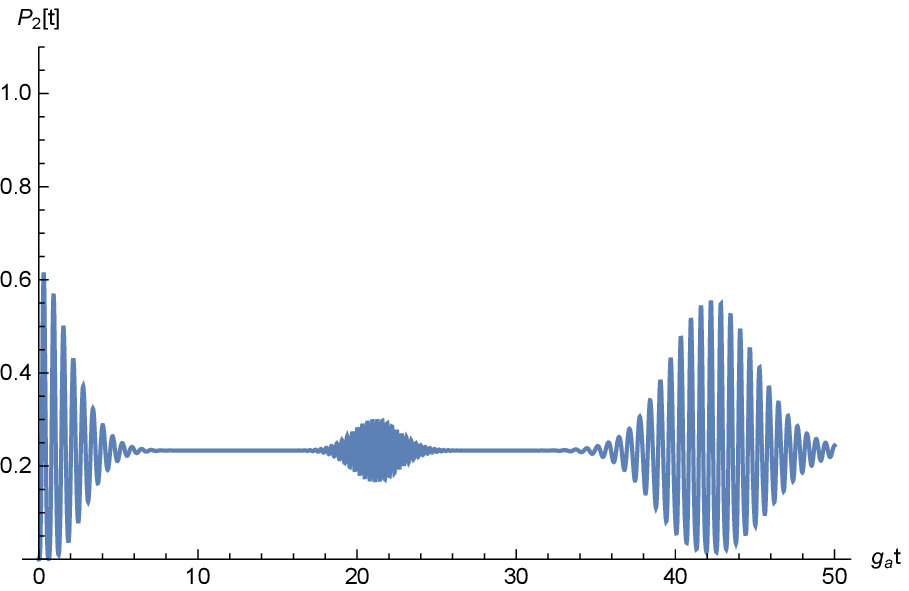}
    \caption{}
  \end{subfigure}
   \begin{subfigure}[b]{0.49\linewidth}
    \includegraphics[width=\linewidth]{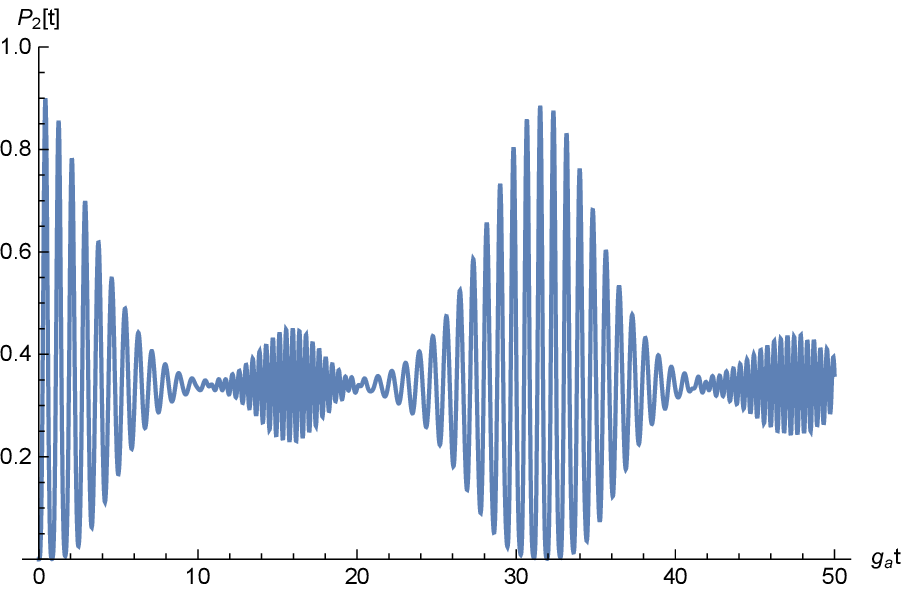}
    \caption{}
  \end{subfigure}
    \captionsetup{width=1\linewidth, font=footnotesize}
  \caption{Time evolution of the atomic occupation probabilities when the initial field is in the two-mode nonlinear CSs with $N=40$, $\mu=1$ and $g_{b}/g_{a}=2$; (a) $\tilde \kappa =0$ and (b) $\tilde \kappa =0.1$}
\end{figure}
As it is evident, for the nonequal coupling strengths, the collapses and revivals of the Rabi oscillations will appear. There are two kinds of the revivals, one with a smaller amplitude and the other with a larger amplitude, which are, respectively, related to one-photon and two-photon processes. Also, by increasing the nonlinearity of the media, the revivals happen in a shorter time periods.

\section{Nonlinear effects on the dynamical properties of the cavity field}

In the present section, we study the field dynamics, particularly the time evolution of the photon distributions, the correlations (or anticorrelations) between the field modes, and the Mandel parameter of each mode.

First, we calculate the time-dependent cross-correlation function (${g^2}\left( t \right) = \frac{{{{\left\langle {{n_a}{n_b}} \right\rangle }_t}}}{{{{\left\langle {{n_a}} \right\rangle }_t}{{\left\langle {{n_b}} \right\rangle }_t}}}$).
For this purpose, we can use the following equations:
\begin{equation}
\begin{array}{l}{\left\langle {{{\hat n}_a}{{\hat n}_b}} \right\rangle _t} = \sum\limits_{{n_a},{n_b}} {{n_a}{n_b}{P_f}({n_a},{n_b};t,\kappa ),} \\{\left\langle {{{\hat n}_a}} \right\rangle _t} = \sum\limits_{{n_a}} {{n_a}P_f^a({n_a};t,\kappa ),} \\{\left\langle {{{\hat n}_b}} \right\rangle _t} = \sum\limits_{{n_b}} {{n_b}P_f^b({n_b};t,\kappa ).} \end{array}\tag{46}
\end{equation}
In order to determine Eq. (46), we should calculate ${P_f}({n_a},{n_b};t,\kappa )$, $P_f^a({n_a};t,\kappa )$ and $P_f^b({n_a};t,\kappa )$.
By using the atom-field state, Eq. (39), the joint photon distribution at time $t$ is given by
\begin{equation}
\begin{array}{l}{P_f}({n_a},{n_b};t,\kappa ) = {p_f}({n_a},{n_b};\kappa){P_0}({n_a},{n_b};t)\\[10pt]
\,\,\,\,\,\,\,\,\,\,\,\,\,\,\,\,\,\,\,\,\,\,\,\ + {p_f}({n_a} + 1,{n_b};t,\kappa ){P_1}({n_a} + 1,{n_b};t)\\[10pt]
\,\,\,\,\,\,\,\,\,\,\,\,\,\,\,\,\,\,\,\,\,\,\,\ + {p_f}({n_a} + 1,{n_b} - 1;\kappa ){P_2}({n_a} + 1,{n_b} - 1;t).\end{array}\tag{47}
\end{equation}\\
By tracing out the appropriate variable, the marginal photon distributions are obtained as\\
\begin{equation}
\begin{array}{l}P_f^a({n_a};t,\kappa ) = \sum\limits_{{n_b}} {{P_f}({n_a},{n_b};t,\kappa )} ,\\P_f^b({n_b};t,\kappa ) = \sum\limits_{{n_a}} {{P_f}({n_a},{n_b};t,\kappa )} .\end{array}\tag{48}
\end{equation}%
For the CK-NCSs of the nonlinear media, Eq. (15), we can write Eqs. (47) and (48), respectively, as:
\begin{equation}
\begin{array}{l}{P_f}({n_a},{n_b};t,\kappa ) = {p_f}({n_a},{n_b};\kappa )\\[10pt]
 \,\,\,\,\,\,\,\,\,\,\,\,\,\,\,\,\,\,\,\,\,\,\,\,\,\,\times{P_0}({n_a},N - {n_a};t){\delta _{{n_b},N - {n_a}}}\\[10pt]
 \,\,\,\,\,\,\,\,\,\,\,\,\,\,\,\,\,\,\,\,\,\,\,\,\,\,\ + {p_f}({n_a} + 1,N - {n_a} - 1;\kappa )\\[10pt]
 \,\,\,\,\,\,\,\,\,\,\,\,\,\,\,\,\,\,\,\,\,\,\,\,\,\,\times{P_1}({n_a} + 1,N - {n_a} - 1;t){\delta _{{n_b},N - {n_a} - 1}}\\[10pt]
 \,\,\,\,\,\,\,\,\,\,\,\,\,\,\,\,\,\,\,\,\,\,\,\,\,\,\ + {p_f}({n_a} + 1,N - {n_a} - 1;\kappa )\\[10pt]
 \,\,\,\,\,\,\,\,\,\,\,\,\,\,\,\,\,\,\,\,\,\,\,\,\,\,\times{P_2}({n_a} + 1,N - {n_a} - 1;t){\delta _{{n_b},N - {n_a}}},\\[10pt]
 P_f^a({n_a};t,\kappa ) = p_f^a(n;\kappa )\\[10pt]
\,\,\,\,\,\,\,\,\,\,\,\,\,\,\,\,\,\,\,\,\,\,\,\,\,\,\,\,\,\times{P_0}(n,N - n;t) \\[10pt]
 \,\,\,\,\,\,\,\,\,\,\,\,\,\,\,\,\,\,\,\,\,\,\,\,\,\,\,\,\ + p_f^a(n + 1;\kappa )\\[10pt]
 \,\,\,\,\,\,\,\,\,\,\,\,\,\,\,\,\,\,\,\,\,\,\,\,\,\,\,\,\,\times[1 - {P_0}(n + 1,N - n - 1;t)],\\[10pt]
 P_f^b({n_b};t,\kappa ) = p_f^b(n - 1;\kappa )\\[10pt]
 \,\,\,\,\,\,\,\,\,\,\,\,\,\,\,\,\,\,\,\,\,\,\,\,\,\,\,\,\,\times{P_2}(N - n + 1,n - 1;t)\\[10pt]
  \,\,\,\,\,\,\,\,\,\,\,\,\,\,\,\,\,\,\,\,\,\,\,\,\,\,\ + p_f^b(n;\kappa )\\[10pt]
  \,\,\,\,\,\,\,\,\,\,\,\,\,\,\,\,\,\,\,\,\,\,\,\,\,\,\,\,\,\times[1 - {P_2}(N - n,n;t)],\end{array}\tag{49}
\end{equation}
where the initial photon distribution ${p_f}({n_a},{n_b};\kappa )$ is given by Eq. (38) and initial distributions $p_f^a(n;\kappa )$ and $p_f^b(n;\kappa )$ are obtained as:
\begin{equation}
\begin{array}{l}p_f^a(n;\kappa ) = \left\langle {{n_a}\left| {{{\hat \rho }_a}} \right|} \right.\left. {{n_a}} \right\rangle\\[10pt]
\,\,\,\,\,\,\,\,\,\,\,\,\,\,\,\,\,\,\,\,\,\,\ = {C^{ - 2}}\left( {\begin{array}{*{20}{c}}N\\n\end{array}} \right){[f(\kappa ,{n_b})!\,]^2}{\left| \mu  \right|^{2n}},\\[10pt]
p_f^b(n;\kappa ) = p_f^a(N - n;\kappa ).\,\,\,\,\,\,\,\,\,\,\,\,\,\,\,\,\,\,\,\,\,\,\,\end{array}\tag{50}
\end{equation}

Fig. 7 displays the cross-correlation function with respect to ${g_a}t$, for different values of the nonlinearity and equal atom-field coupling constants. As it is seen, the cross-correlation function have a periodic behavior and oscillates between its initial value and some lower values. Also, the effect of the interaction is to increase the anticorrelation between the modes, for all values of the nonlinearity. This increase in anticorrelation is derived from the periodic annihilation of a photon from the mode $a$ and the creation of the same photon into the mode $b$. Furthermore, it is obvious that by increasing the nonlinearity $\tilde \kappa $, the anticorrelation between the modes is decreased.

In Fig. 8, we have plotted the cross-correlation function with respect to ${g_a}t$ , for different values of the nonlinearity $\tilde \kappa $ and for nonequal atom-field coupling constants. Similar to the case of equal atom-field coupling constant, by increasing $\tilde \kappa $, the anticorrelation between the modes is decreased. In addition, the collapse and revivals of the Rabi oscillations will appear similar to the atomic occupation probabilities. However, these revivals occur in pairs unlike the atomic occupation probabilities. This difference is related to the particular expression $\left\langle {{n_a}} \right.{\left. {{n_b}} \right\rangle _t}$, which is relayed on the occupation properties of two atomic levels.
\begin{figure}[h]
 \centering
  \includegraphics[width=1\linewidth]{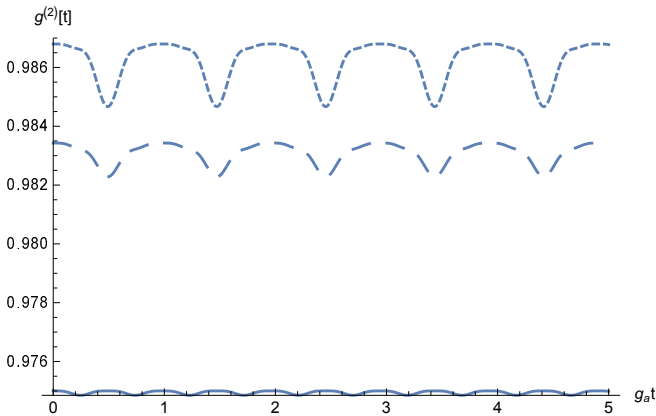}
    \captionsetup{width=1\linewidth, font=footnotesize}
  \caption{Cross-correlation function versus ${g_a}t$ for $\tilde \kappa =0$ (solid line), $\tilde \kappa =0.01$ (dashed line) and $\tilde \kappa =0.1$ (dotted line) with $N=40$, $\mu=1$ and $g_{b}/g_{a}=1$.}
\end{figure}

\begin{figure}[h]
 \centering
  \includegraphics[width=1\linewidth]{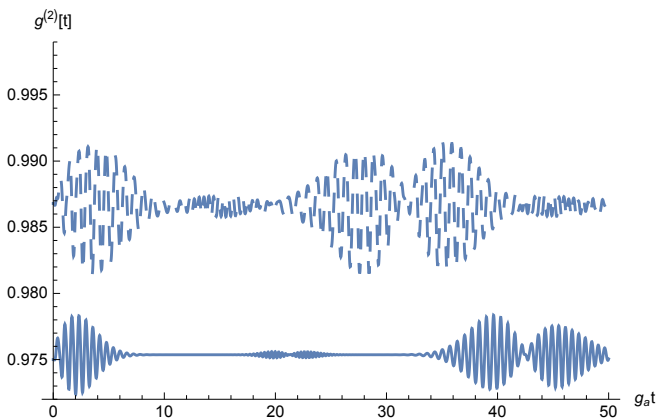}
    \captionsetup{width=1\linewidth, font=footnotesize}
  \caption{Cross-correlation function versus ${g_a}t$ for $\tilde \kappa =0$ (solid line) and $\tilde \kappa =0.1$ (dashed line), $N=40$, $\mu=1$ and $g_{b}/g_{a}=2$.}
 \end{figure}

In order to determine the photon statistics of the two modes, we consider time-dependent Mandel parameter as~\cite{vogel1995nonlinear}
\begin{equation}
{Q_i}\left( t \right) = \frac{{\left( {\Delta {n_i}} \right)_t^2 - {{\left\langle {{{\hat n}_i}} \right\rangle }_t}}}{{{{\left\langle {{{\hat n}_i}} \right\rangle }_t}}},\,\,\,\,\,\,\,\,i = a,b. \tag{51}
\end{equation}

In Figs. 9 and 10, respectively, we have plotted the Mandel parameters of the two modes versus ${g_a}t$ with $N = 10$, $\mu  = 1$, ${{{g_b}} \mathord{\left/
 {\vphantom {{{g_b}} {{g_a} = 1}}} \right.
 \kern-\nulldelimiterspace} {{g_a} = 1}}$, and for different values of $\tilde \kappa $. As it is seen, both modes have periodic sub-Poissonian statistics. Furthermore, by increasing the $\tilde \kappa $, the photon-counting statistics of the first (second) mode of the cavity tends to sub-Poissonian (Poissonian) more rapidly. In other words, the first (second) mode of the cavity on the nonlinear medium shows more (less) nonclassical properties than the first (second) mode of the medium without nonlinearity.
\begin{figure}[htb]
  \centering
  \begin{subfigure}[b]{0.49\linewidth}
    \includegraphics[width=\linewidth]{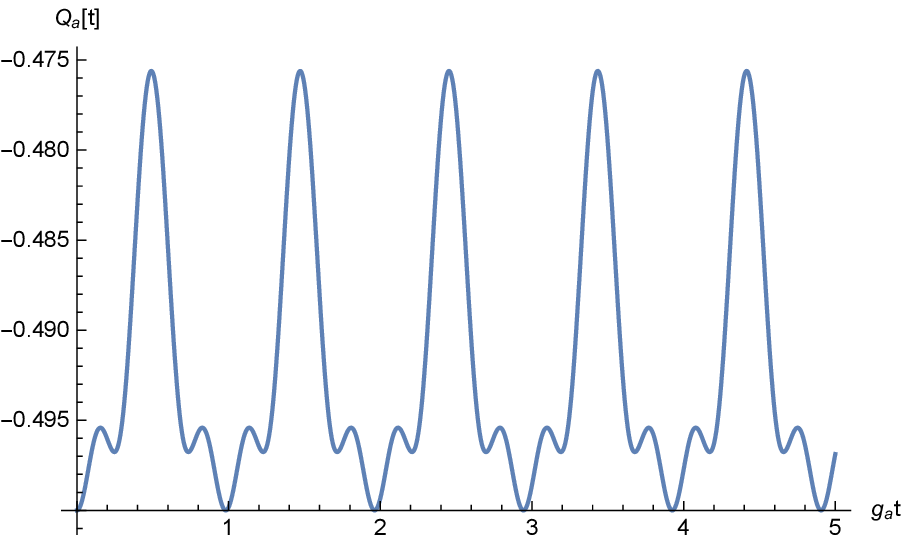}
    \caption{}
  \end{subfigure}
  \begin{subfigure}[b]{0.49\linewidth}
    \includegraphics[width=\linewidth]{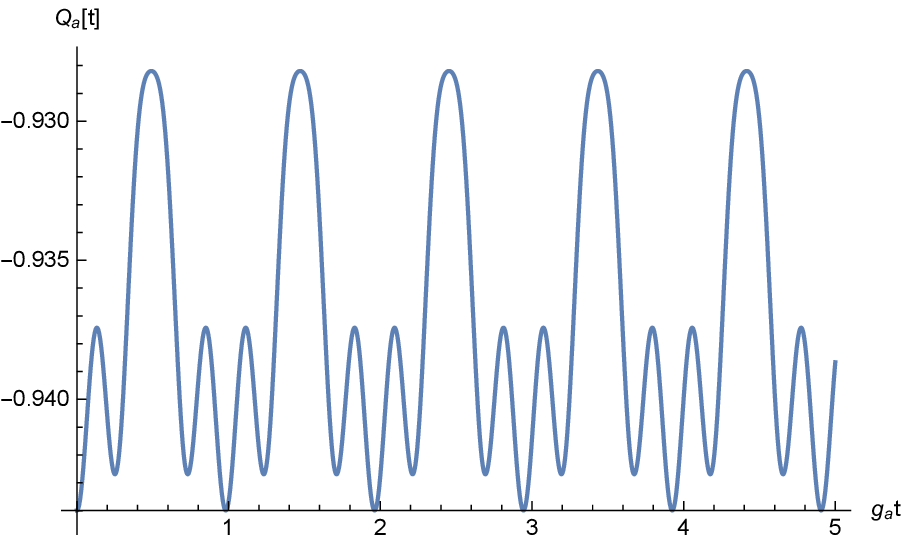}
     \caption{}
  \end{subfigure}
    \captionsetup{width=1\linewidth, font=footnotesize}
  \caption{Mandel parameter of the first mode of the field versus ${g_a}t$ for (a) $\tilde \kappa =0$ and (b) $\tilde \kappa =0.1$, with $N=40$, $\mu=1$ and $g_{b}/g_{a}=1$.}

\end{figure}
\begin{figure}[htb]
  \centering
  \begin{subfigure}[b]{0.49\linewidth}
    \includegraphics[width=\linewidth]{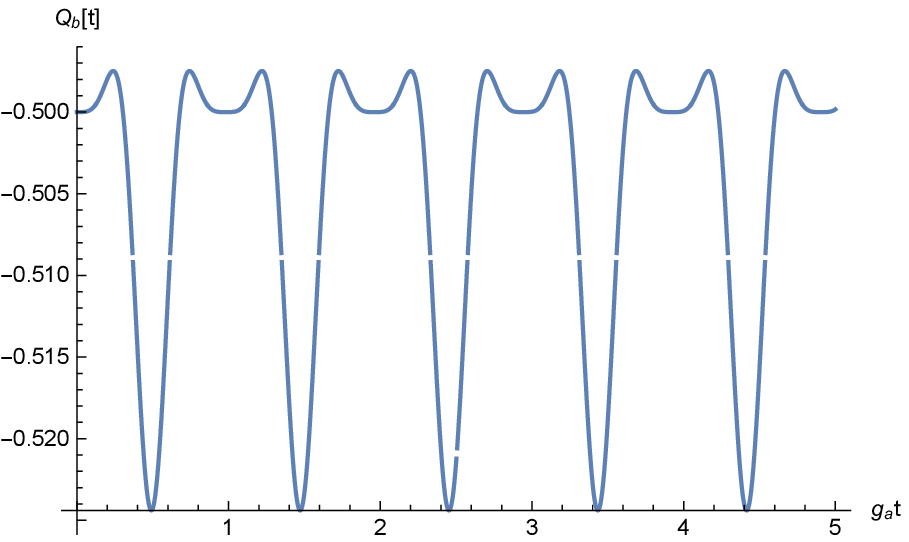}
    \caption{}
  \end{subfigure}
  \begin{subfigure}[b]{0.49\linewidth}
    \includegraphics[width=\linewidth]{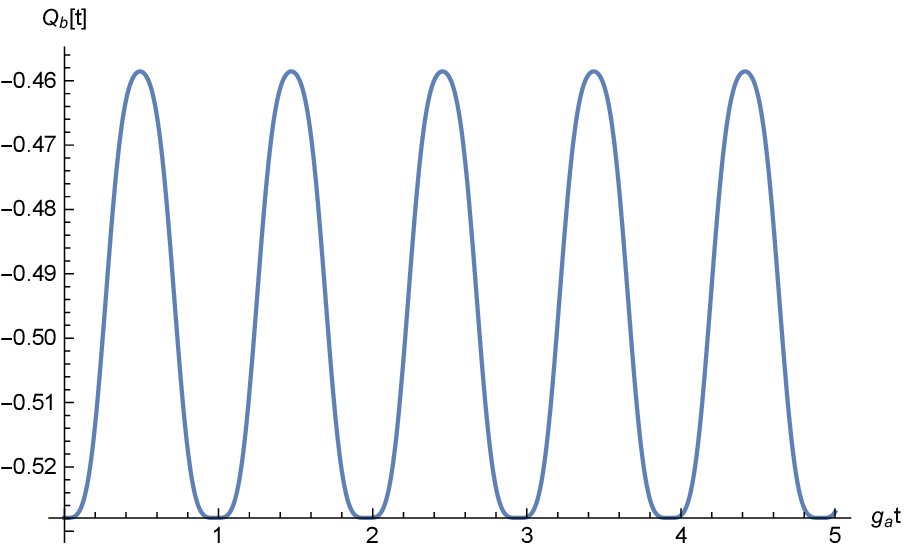}
    \caption{}
  \end{subfigure}
    \captionsetup{width=1\linewidth, font=footnotesize}
  \caption{Mandel parameter of the second mode of the field versus ${g_a}t$ for (a) $\tilde \kappa =0$ and (b) $\tilde \kappa =0.1$, with $N=40$, $\mu=1$ and $g_{b}/g_{a}=1$.}

\end{figure}

Figs. 11 and 12 display the Mandel parameters of the two modes as a function of ${g_a}t$ for different values of the $\kappa $ with nonequal atomic-field coupling strengths
${{{g_b}} \mathord{\left/
 {\vphantom {{{g_b}} {{g_a} = 2}}} \right.
 \kern-\nulldelimiterspace} {{g_a} = 2}}$.
Similar to the time-evolution of the cross-correlation function, the collapse-revival phenomenon occurs in the time evolution of the Mandel parameters and the revivals occur in pairs. Also, by increasing the $\tilde \kappa $, revivals occur in shorter time periods.
\begin{figure}[htb]
  \centering
  \begin{subfigure}[b]{0.49\linewidth}
    \includegraphics[width=\linewidth]{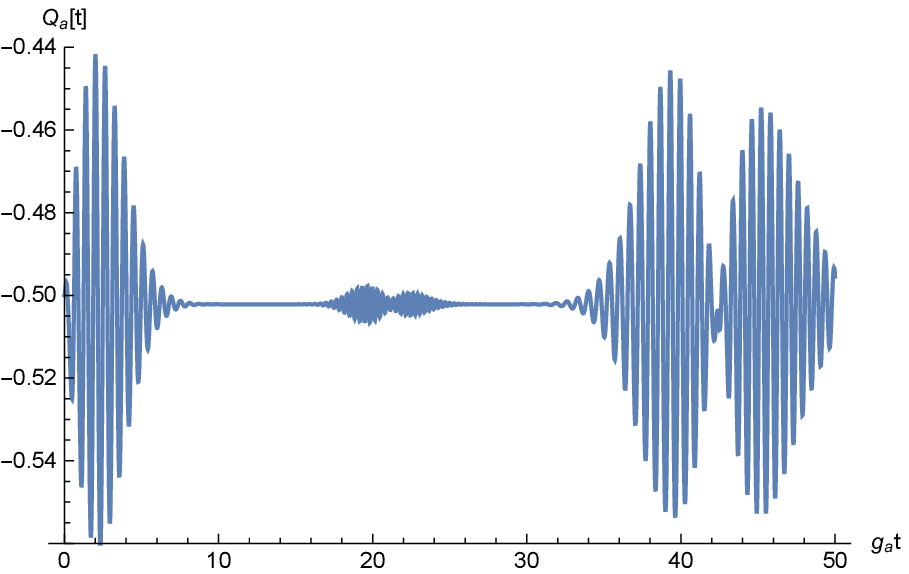}
    \caption{}
  \end{subfigure}
  \begin{subfigure}[b]{0.49\linewidth}
    \includegraphics[width=\linewidth]{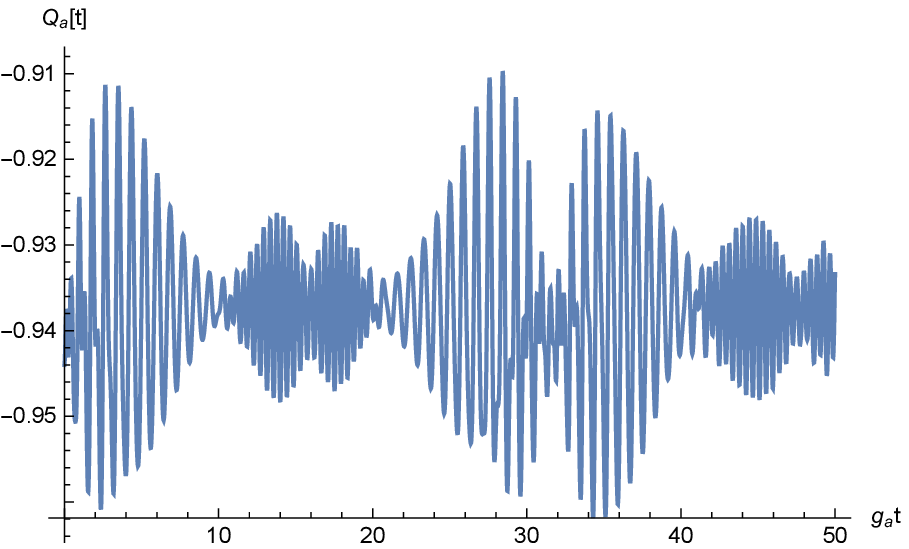}
    \caption{}
  \end{subfigure}
    \captionsetup{width=1\linewidth, font=footnotesize}
  \caption{Mandel parameter of the first mode of the field versus ${g_a}t$ for (a) $\tilde \kappa =0$ and (b) $\tilde \kappa =0.1$, with $N=40$, $\mu=1$ and $g_{b}/g_{a}=2$.}
\end{figure}

\begin{figure}[htb]
  \centering
  \begin{subfigure}[b]{0.49\linewidth}
    \includegraphics[width=\linewidth]{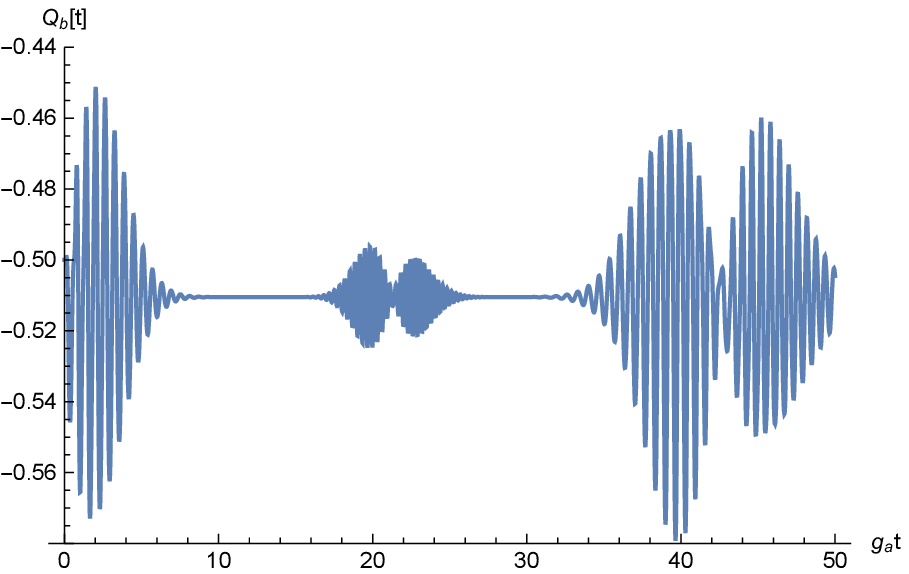}
    \caption{}
  \end{subfigure}
  \begin{subfigure}[b]{0.49\linewidth}
    \includegraphics[width=\linewidth]{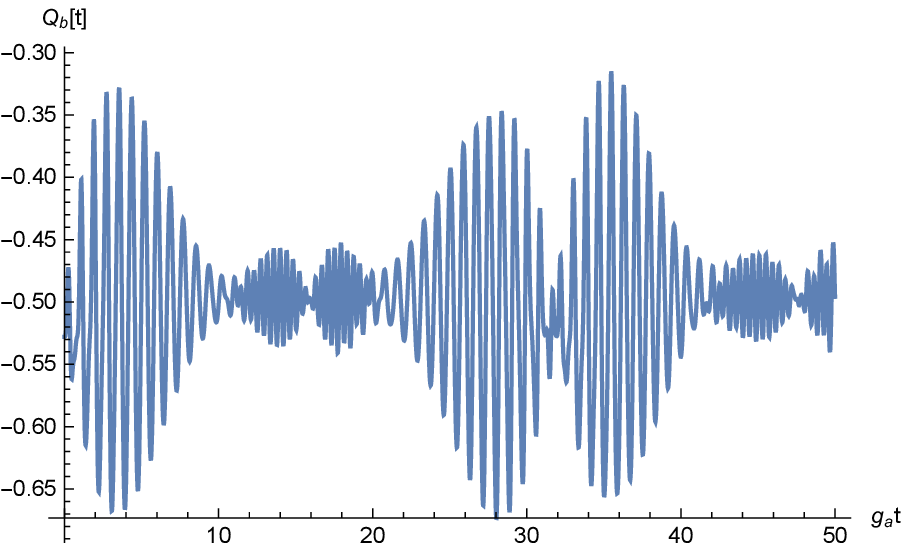}
    \caption{}
  \end{subfigure}
    \captionsetup{width=1\linewidth, font=footnotesize}
  \caption{Mandel parameter of the second mode of the field versus ${g_a}t$ for (a) $\tilde \kappa =0$ and (b) $\tilde \kappa =0.1$, with $N=40$, $\mu=1$ and $g_{b}/g_{a}=2$.}

\end{figure}

As another important nonclassical property, we investigate the squeezing. In order to obtain the two-mode squeezing of the system, we consider two-mode quardrature operators, which have been defined as~\cite{gerry2005introductory}
\begin{equation}
\begin{array}{l}{{\hat X}_1} = \frac{1}{{2\sqrt 2 }}\left( {\hat a + {{\hat a}^\dag } + \hat b + {{\hat b}^\dag }} \right),\\{{\hat X}_2} = \frac{1}{{2i\sqrt 2 }}\left( {\hat a - {{\hat a}^\dag } + \hat b - {{\hat b}^\dag }} \right).\end{array}\tag{52}
\end{equation}
These definitions lead to the commutation relation $[{\hat X_1},{\hat X_2}] = {\raise0.7ex\hbox{$i$} \!\mathord{\left/
 {\vphantom {i 2}}\right.\kern-\nulldelimiterspace}
\!\lower0.7ex\hbox{$2$}}$. The commutation relation for two-mode leads to the following uncertainty relation,
\begin{equation}
{(\Delta {\hat X_1})^2}{(\Delta {\hat X_2})^2} \ge \frac{1}{{16}}.\tag{53}
\end{equation}
A state is squeezed in ${\hat X_1}\,({\hat X_2})$ if ${(\Delta {\hat X_1})^2} < 0.25\,(\,{(\Delta {\hat X_2})^2} < 0.25)$, or equivalently ${S_{{X_{1(2)}}}} = 4{(\Delta {\hat X_{1(2)}})^2} - 1< 0(>0)$.\\

In Figs. 13(a) and 13(b), respectively, we have plotted ${S_{{X_1}}}$ and ${S_{{X_2}}}$ with respect ${g_a}t$ for $N = 10$ and different values of $\tilde \kappa $. These Figs. clearly show that by increasing $\tilde \kappa $, the degree of quadrature squeezing is enhanced. In summary, we can result that by going from linear to nonlinear medium, we encounter with more nonclassical properties.
\begin{figure}[htb]
  \centering
  \begin{subfigure}[b]{0.49\linewidth}
    \includegraphics[width=\linewidth]{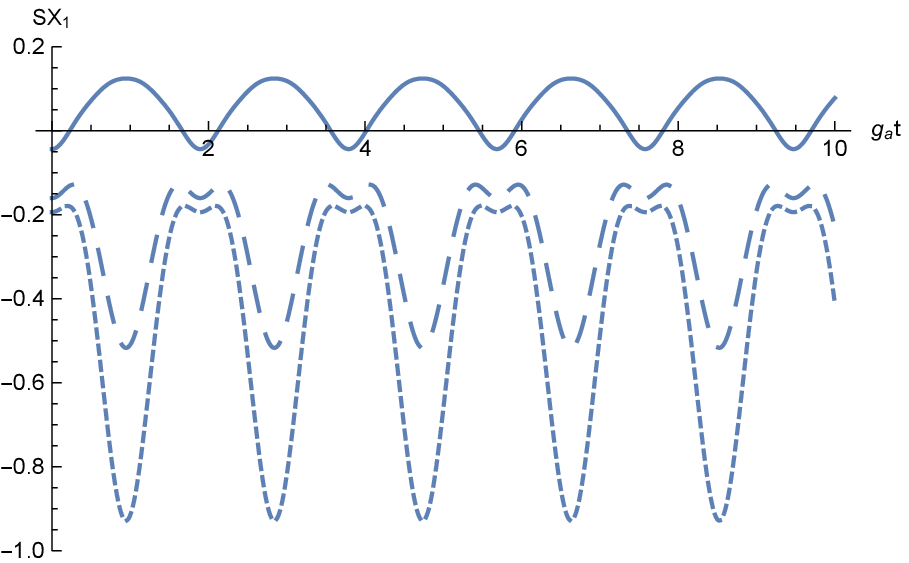}
    \caption{}
  \end{subfigure}
  \begin{subfigure}[b]{0.49\linewidth}
    \includegraphics[width=\linewidth]{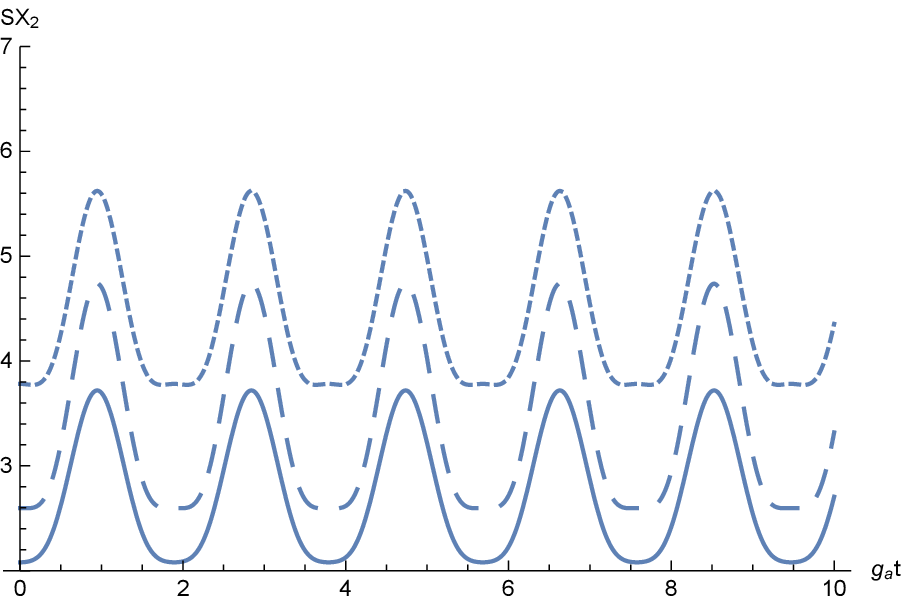}
    \caption{}
  \end{subfigure}
    \captionsetup{width=1\linewidth, font=footnotesize}
  \caption{$S_{X_{1}}$ and $S_{X_{2}}$ versus ${g_a}t$ for $\tilde \kappa =0$ (solid line), $\tilde \kappa =0.1$ (dashed line) and $\tilde \kappa =0.5$ (dotted line), with $N=10$.}

\end{figure}

To study the dynamics of the entanglement, one must choose an entanglement measure. For the present case, we use von Neumann entropy. It is worth nothing that the von Neumann entropy is a measure of entanglement for pure bipartite states~\cite{phoenix1988fluctuations}. The von Neumann entropy is given by:
\begin{equation}
S(\hat \rho ) =  - Tr[\hat \rho \ln \hat \rho ],\tag{54}
\end{equation}%
where $\rho $ is the density operator for a given quantum system. If $\rho $ describes a pure state (mixed state), then $S = 0$ ($S \ne 0$). Furthermore, if ${S_A}$ and ${S_F}$ denote the entropies of two interacting systems and $S$ denote entropy of the composite system, Araki and Leib showed that these entropies satisfy the following inequality~\cite{araki2002entropy}\\
\begin{equation}
\left| {{S_A} - {S_F}} \right| \le S \le {S_A} + {S_F}.\tag{55}
\end{equation}
For our model, the entropy of the atom (field) is defined through the corresponding reduced density operator as,\\
\begin{equation}
 {S_{A(F)}}(\hat \rho ) =  - T{r_{A(F)}}[{\hat \rho _{A(F)}}\ln {\hat \rho _{A(F)}}].\tag{56}
\end{equation}
Here, we use this entropy as a measure of the degree of entanglement between the field and the atom. We assume that the system starts from a pure state as stated in Eq. (37). Consequently, its entropy $S$ vanishes. Therefore, inequality (55) implies that ${S_A} = {S_F}$. Since the trace is invariant under a similarity transformation, we can go to bases in which the atomic density matrix is diagonal and write Eq. (56) as
\begin{equation}
 {S_A} = {S_F} =  - \sum\limits_{i = 1}^3 {{\lambda _i}\ln {\lambda _i}} ,\tag{57}
\end{equation}%
where ${\lambda _i}$ are the eigenvalues for the atomic density matrix ${\rho _A}(t)$ . Now, we obtain the final state of the system $\rho (t)$ at any time $t > 0$ as
\begin{equation}
\rho (t) = \left| {\psi (t)} \right\rangle \left\langle {\psi (t)} \right|.\tag{58}
\end{equation}
The elements of the atomic reduced density matrix ${\rho _A}(t)$ can be expressed as
\begin{equation}
\begin{array}{l}{\rho _{11}} = \sum\limits_{{n_a},{n_b}} {{{\left| {{A_{{n_a},{n_b}}}(\kappa )} \right|}^2}} {\left| {{C_0}({n_a},{n_b};t)} \right|^2},\\[10pt]
{\rho _{22}} = \sum\limits_{{n_a},{n_b}} {{{\left| {{A_{{n_a},{n_b}}}(\kappa )} \right|}^2}} {\left| {{C_1}({n_a} + 1,{n_b};t)} \right|^2},\\[10pt]
{\rho _{33}} = \sum\limits_{{n_a},{n_b}} {{{\left| {{A_{{n_a},{n_b}}}(\kappa )} \right|}^2}} {\left| {{C_2}({n_a} + 1,{n_b} - 1;t)} \right|^2},\\[10pt]
{\rho _{12}} = \sum\limits_{{n_a},{n_b}} {{A_{{n_a},{n_b}}}(\kappa ){A_{{n_a} + 1,{n_b}}}(\kappa )} {C_0}({n_a},{n_b};t)\\[10pt]
\,\,\,\,\,\,\,\,\,\,\,\,\,\,\,\ \times{C_1}^*({n_a} + 1,{n_b};t),\\[10pt]
{\rho _{21}} = \sum\limits_{{n_a},{n_b}} {{A_{{n_a},{n_b}}}(\kappa ){A_{{n_a} + 1,{n_b}}}(\kappa )} C_0^*({n_a},{n_b};t)\\[10pt]
\,\,\,\,\,\,\,\,\,\,\,\,\,\,\,\ \times{C_1}({n_a} + 1,{n_b};t),\\[10pt]
{\rho _{13}} = \sum\limits_{{n_a},{n_b}} {{A_{{n_a},{n_b}}}(\kappa ){A_{{n_a} + 1,{n_b} - 1}}(\kappa )} {C_0}({n_a},{n_b};t)\\[10pt]
\,\,\,\,\,\,\,\,\,\,\,\,\,\,\,\ \times{C_2}^*({n_a} + 1,{n_b} - 1;t),\\[10pt]
{\rho _{31}} = \sum\limits_{{n_a},{n_b}} {{A_{{n_a},{n_b}}}(\kappa ){A_{{n_a} + 1,{n_b} - 1}}(\kappa )} {C_0}^*({n_a},{n_b};t)\\[10pt]
\,\,\,\,\,\,\,\,\,\,\,\,\,\,\,\ \times{C_2}({n_a} + 1,{n_b} - 1;t),\\[10pt]
{\rho _{23}} = \sum\limits_{{n_a},{n_b}} {{A_{{n_a},{n_b}}}(\kappa ){A_{{n_a},{n_b} + 1}}(\kappa )} {C_1}({n_a} + 1,{n_b};t)\\[10pt]
\,\,\,\,\,\,\,\,\,\,\,\,\,\,\,\ \times{C_2}^*({n_a} + 1,{n_b} - 1;t),\\[10pt]
{\rho _{32}} = \sum\limits_{{n_a},{n_b}} {{A_{{n_a},{n_b}}}(\kappa ){A_{{n_a},{n_b} + 1}}(\kappa )} C_1^*({n_a} + 1,{n_b};t)\\[10pt]
\,\,\,\,\,\,\,\,\,\,\,\,\,\,\,\ \times{C_2}({n_a} + 1,{n_b} - 1;t).\end{array}\tag{59}
\end{equation}
Furthermore, the eigenvalues ${\lambda _i}$ of the atomic density matrix ${\rho _A}(t)$ are the roots of the cubic equation
\begin{equation}
{\lambda ^3} + {b_0}{\lambda ^2} + {b_1}\lambda  + {b_2} = 0,\tag{60}
\end{equation}
where the coefficients ${b_0}$ ,${b_1}$ and ${b_2}$ are given by
\begin{equation}
\begin{array}{l}{b_0} =  - ({\rho _{11}} + {\rho _{22}} + {\rho _{33}}),\\
{b_1} = {\rho _{11}}{\rho _{22}} + {\rho _{11}}{\rho _{33}} + {\rho _{22}}{\rho _{33}} \\
\,\,\,\,\,\,\,\,\ - {\rho _{12}}{\rho _{21}} - {\rho _{13}}{\rho _{31}} - {\rho _{23}}{\rho _{32}},\\
{b_2} = {\rho _{11}}{\rho _{23}}{\rho _{32}} + {\rho _{22}}{\rho _{13}}{\rho _{31}} + {\rho _{33}}{\rho _{12}}{\rho _{21}}\\
 \,\,\,\,\,\,\,\,\ - {\rho _{11}}{\rho _{22}}{\rho _{33}} - {\rho _{13}}{\rho _{21}}{\rho _{32}} - {\rho _{12}}{\rho _{23}}{\rho _{31}}.\end{array}\tag{61}
\end{equation}
Therefore, the eigenvalues ${\lambda _i}(i = 1, 2, 3)$ are obtained as:
\begin{equation}
\begin{array}{l}{\lambda _i}(t) =  - \frac{{{b_0}}}{3} + \frac{2}{3}\sqrt {b_0^2 - 3{b_2}} \cos \left[ {\alpha  + \frac{2}{3}(i - 1)\pi } \right],\end{array}\tag{62}
\end{equation}
where,
\begin{equation}
 \alpha  = \frac{1}{3}co{s^{ - 1}}\left( {\frac{{9{b_0}{b_1} - 2b_0^3 - 27{b_2}}}{{2{{(b_0^2 - 3{b_2})}^{3/2}}}}} \right).\tag{63}
\end{equation}%
Now, to obtain the entanglement between the atom and the field, we discuss the dynamical behavior of the entropy of the sub-systems, according to Eq. (57).

Figs. 14(a) and 14(b) display the time evolution of the field entropies with respect ${g_a}t$ for $N = 10$. As it is seen, for equal atomic-field coupling strengths ${{{g_b}} \mathord{\left/
 {\vphantom {{{g_b}} {{g_a} = 1}}} \right.
 \kern-\nulldelimiterspace} {{g_a} = 1}}$, the time evolution of the von Neumann entropy  has a periodic behavior. Also, for the nonequal coupling strengths, a chaotic behaviour for the time evolution of the field entropy is observed. Also, comparing the Figs. 14(a) and 14(b) indicates that by choosing the nonequal coupling strengths, the maximum value of the quantum field entropy is increased.
\begin{figure}[htb]
  \centering
  \begin{subfigure}[b]{0.49\linewidth}
    \includegraphics[width=\linewidth]{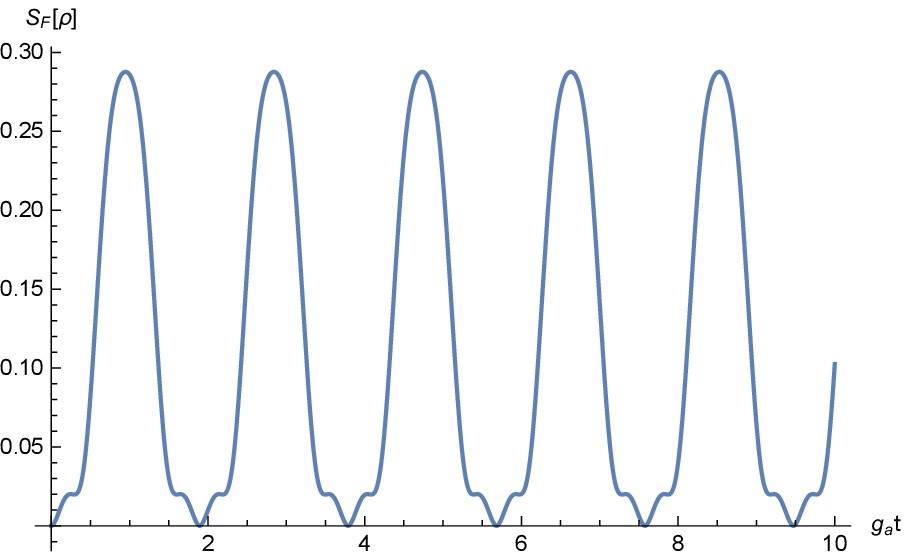}
    \caption{}
  \end{subfigure}
  \begin{subfigure}[b]{0.49\linewidth}
    \includegraphics[width=\linewidth]{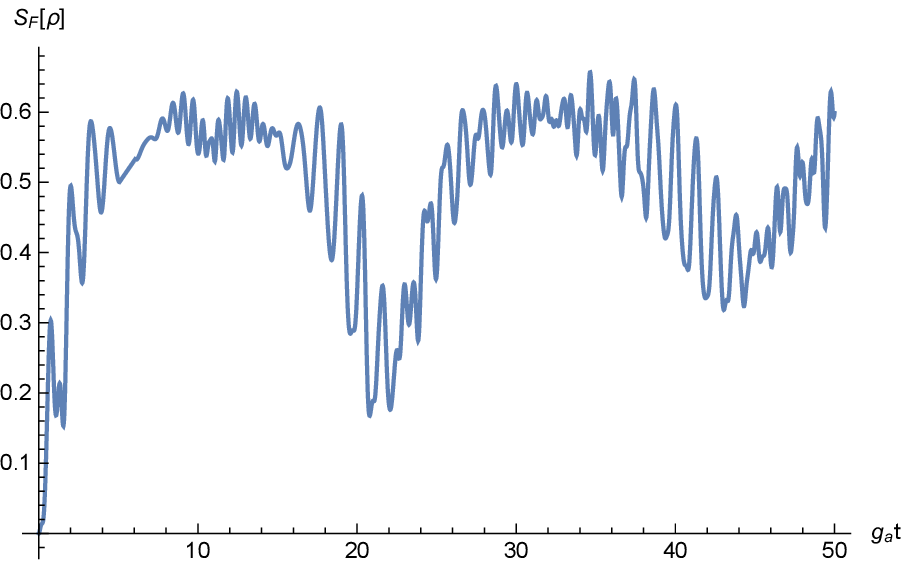}
    \caption{}
  \end{subfigure}
    \captionsetup{width=1\linewidth, font=footnotesize}
  \caption{The time evolution of the von Neumann entropy versus ${g_a}t$ with $N=10$, $\tilde \kappa = 0$, $\mu=1$ for (a) $g_{b}/g_{a}=1$ and (b) $g_{b}/g_{a}=2$.}

\end{figure}

\begin{figure}[htb]
  \centering
  \begin{subfigure}[b]{0.49\linewidth}
    \includegraphics[width=\linewidth]{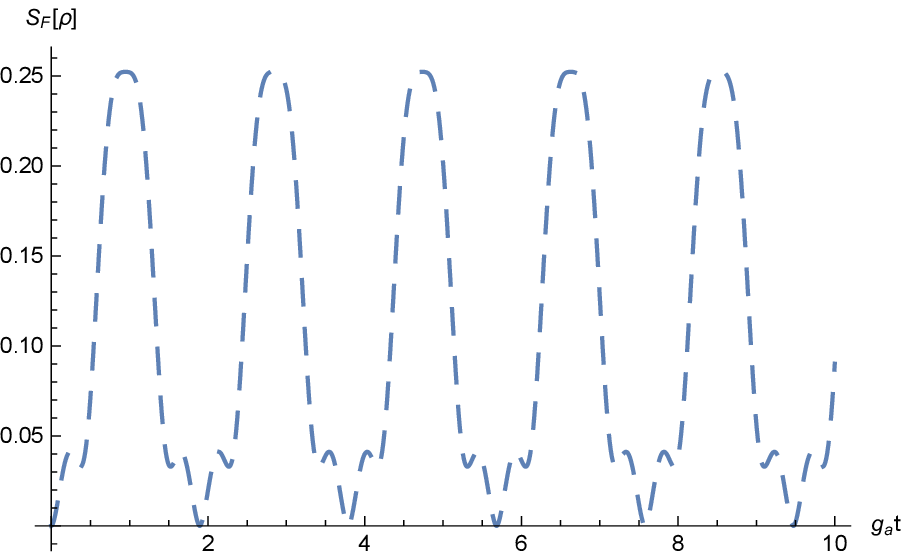}
    \caption{}
  \end{subfigure}
  \begin{subfigure}[b]{0.49\linewidth}
    \includegraphics[width=\linewidth]{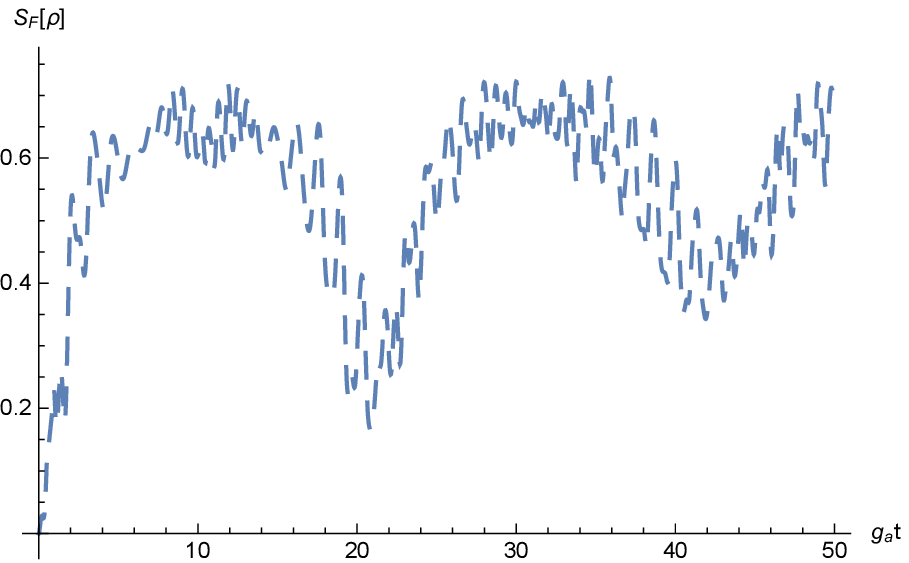}
    \caption{}
  \end{subfigure}
    \captionsetup{width=1\linewidth, font=footnotesize}
  \caption{The time evolution of the von Neumann entropy versus ${g_a}t$ with $N=10$, $\tilde \kappa = 0.1$, $\mu=1$ for (a) $g_{b}/g_{a}=1$ and (b) $g_b/g_a=2$.}

\end{figure}

\begin{figure}[htb]
  \centering
  \begin{subfigure}[b]{0.49\linewidth}
    \includegraphics[width=\linewidth]{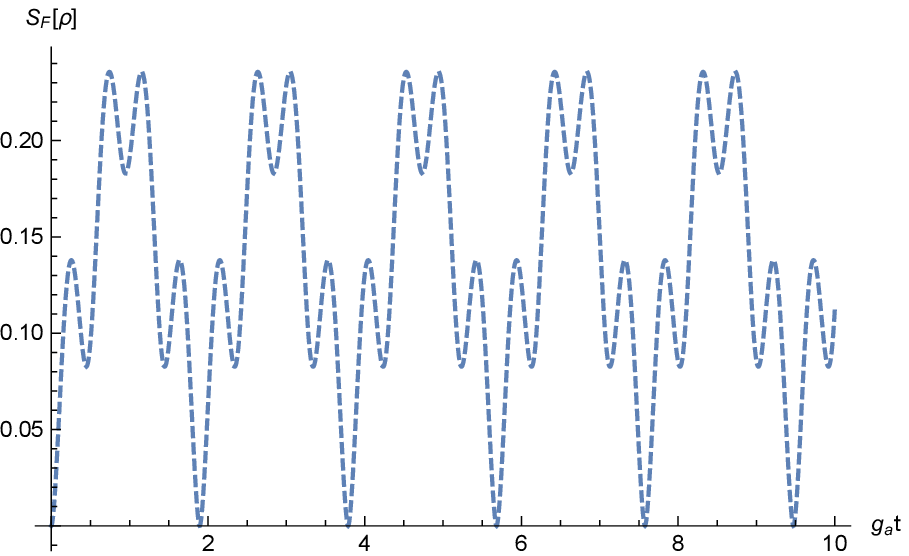}
    \caption{}
  \end{subfigure}
  \begin{subfigure}[b]{0.49\linewidth}
    \includegraphics[width=\linewidth]{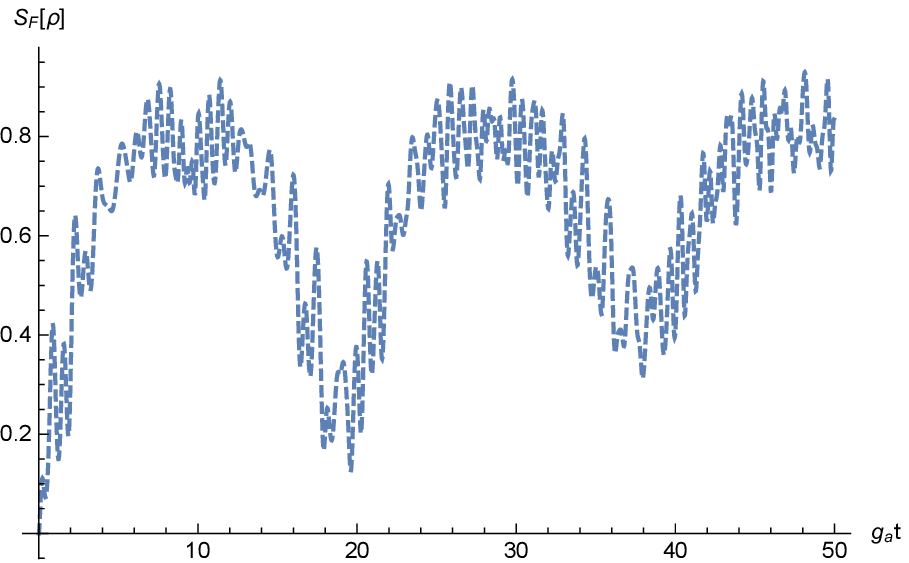}
    \caption{}
  \end{subfigure}
    \captionsetup{width=1\linewidth, font=footnotesize}
  \caption{The time evolution of the von Neumann entropy versus ${g_a}t$ with $N = 10$, $\tilde \kappa  = 1$, $\mu  = 1$ for (a) $g_{b}/g_{a}=1$ and (b) $g_b/g_a=2$.}

\end{figure}

In Figs. 15 and 16 we have plotted the time evolution of the von Neumann entropy versus ${g_a}t$ for $\tilde \kappa  = 0.1$ and $\tilde \kappa  = 1$, respectively. In these cases, similar to the Fig. 14, the time evolution of the entropy for equal atomic-field coupling strengths ${{{g_b}} \mathord{\left/
 {\vphantom {{{g_b}} {{g_a} = 1}}} \right.
 \kern-\nulldelimiterspace} {{g_a} = 1}}$ has a periodic behavior and by entering the effect of atomic-field coupling strengths, a chaotic behaviour for the time evolution of the field entropy is revealed. Also, by including the effect of nonequal coupling strengths, the maximum value of the entropy is increased.

Furthermore, by comparing Figs. 14, 15 and 16, it is observed that for the equal atomic-field coupling strengths, entanglement is decreased by increasing the $\tilde \kappa $ ( see Fig. 17(a)). On the other hand, the amount of entanglement for nonequal atomic-field coupling strengths is increased by increasing $\tilde \kappa $ (see Fig. 17(b)). In summary, the numerical results of the von Neumann entropy showed that the existence of the cross-Kerr medium decreases the maximum amount of entanglement between atomic and field subsystems for equal atomic-field coupling strengths, while nonequal the atomic-field coupling strengths has different effect on the maximum value of the entanglement.
\begin{figure}[htb]
  \centering
  \begin{subfigure}[b]{0.49\linewidth}
    \includegraphics[width=\linewidth]{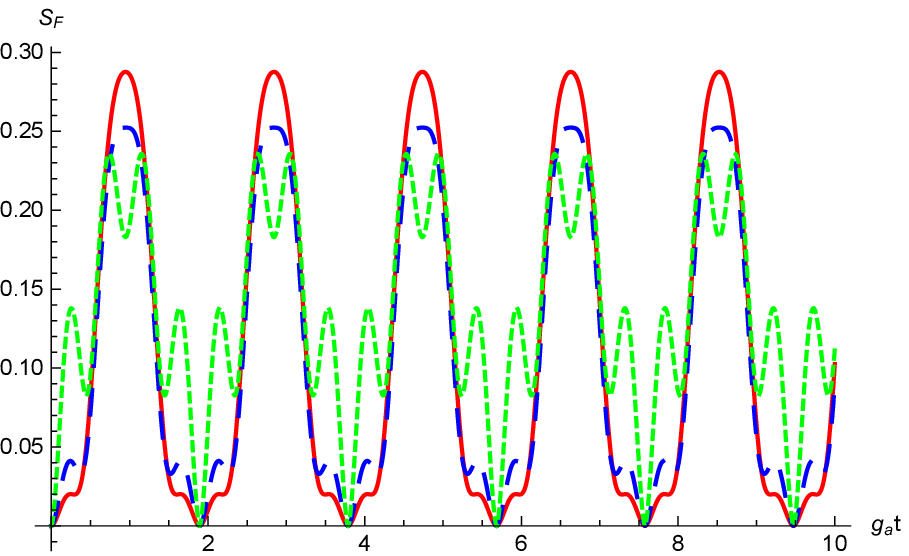}
    \caption{}
  \end{subfigure}
  \begin{subfigure}[b]{0.49\linewidth}
    \includegraphics[width=\linewidth]{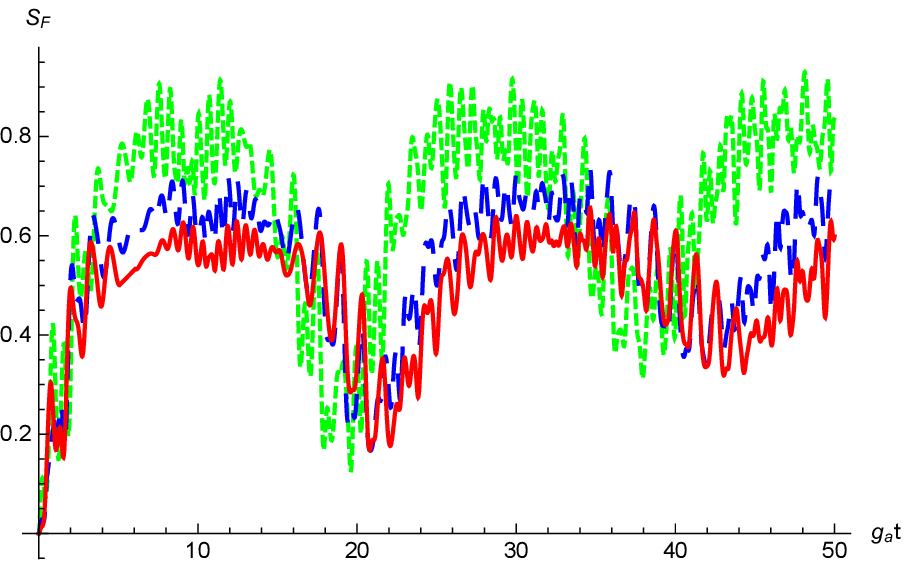}
    \caption{}
  \end{subfigure}
    \captionsetup{width=1\linewidth, font=footnotesize}
\caption{The time evolution of the von Neumann entropy versus ${g_a}t$ for $\tilde \kappa  = 0$ (solid red line), $\tilde \kappa  = 0.1$ (dashed blue line) and $\tilde \kappa  = 1$ (dotted green line), with $N = 10$, $\mu  = 1$ for (a) $g_{b}/g_{a}=1$ and (b) $g_b/g_a=2$.}

\end{figure}
\section{Summary and concluding remarks}
In this paper, at first, we have searched for a relation between deformation function of the f-deformed oscillator algebra and a two-mode cross-Kerr nonlinear optical algebra. We have found that it is possible to describe the cross-Kerr media as a deformed two-dimensional harmonic oscillator algebra, as well as a deformed $su(2)$ algebra. Then, we have constructed the NCSs corresponding to this cross-Kerr system and studied their quantum statistical properties. The results show that anticorrelated two modes of the constructed CSs have sub-Poissonain statistics. Furthermore, the effect nonlinearity of the media on the nonclassical properties of two modes is clarified.
In the following, we proposed a theoretical scheme to investigate the nonlinear effects of the nonlinear media via nonlinear CSs approach. For this purpose, we have considered the interaction of the two-mode CK-NCSs with a  $\Lambda $-type three-level atom. We have found that the atomic occupation probabilities of the various atomic levels for equal atomic-field coupling strengths have a periodic behavior and by increasing  $\tilde \kappa $, the one-photon transition is increased. For the nonequal coupling strengths, the collapses and revivals of the Rabi oscillations observed and by increasing the nonlinearity of the media, the revivals occurred in shorter time periods. By studying the field dynamics, such as cross-correlation function, we have shown that by increasing the nonlinearity of the media, the anticorrelation between the modes is decreased. To investigate the effects of the nonlinearity on the photon counting statistics of the two modes of the cavity field, we have examined the behavior of the Mandel parameters of the two modes. Both and found that both modes had periodic sub-Poissonian statistics. Furthermore, by increasing the nonlinearity of the media, the photon-counting statistics of the first (second) mode tended to sub-Poissonian (Poissonian). By studying squeezing phenomenon, we have shown that by going from linear medium to nonlinear medium, CK-NCSs exhibited more nonclassical properties. In the continuation of our research, we have used the von Neumann entropy as a measure of the degree of entanglement between the field and the atom of the system under consideration. We have shown that the existence of the cross-Kerr medium, decreased the maximum amount of entanglement between atomic and field subsystems for equal atomic-field coupling strengths, while nonequal the atomic-field coupling strengths had different effect on the maximum value of the entropy.


\end{document}